\documentclass[10pt]{article}
\usepackage{appendix}
\usepackage{setspace}
\usepackage[font=small,  margin=2cm]{caption}
\usepackage[tbtags]{amsmath}
\usepackage{amsthm,amssymb,amsfonts}
\usepackage[numbers]{natbib}
\usepackage{multirow}
\usepackage{lscape}
\usepackage{subfigure}
\usepackage{makecell}
\usepackage{exscale}
\usepackage{booktabs}
\usepackage{array}
\usepackage{fullpage}
\usepackage{url}
\usepackage{algorithm}
\usepackage{algpseudocode}
\usepackage{bm}
\usepackage{smile}
\usepackage{mathtools}
\usepackage{wrapfig}
\usepackage{lipsum}
\usepackage{mathrsfs}
\usepackage{relsize}
\usepackage{dsfont}
\usepackage{multirow}
\usepackage{apalike}
\usepackage{chngcntr}
\usepackage[usenames,dvipsnames,svgnames,table]{xcolor}
\usepackage[colorlinks=true,
linkcolor=blue,
urlcolor=blue,
citecolor=blue]{hyperref}

\def\reals{{\mathbb R}}

\def\hbbeta{\widehat \bbeta}

\def\hepsilon{\widehat \epsilon}

\def\P{{\mathbb P}}
\def\E{{\mathbb E}}

\newcommand{\bop}{\bo_{p}}
\newcommand{\bOp}{\bO_{p}}

\newcommand{\bmax}{\mathop{\mathrm{max}}}

\numberwithin{equation}{section}
\numberwithin{theorem}{section}
\numberwithin{corollary}{section}
\counterwithout{asmp}{section}
\numberwithin{definition}{section}

\begin{document}

\title{\LARGE High-dimensional Two-sample Precision Matrices Test:  An \\ Adaptive Approach through Multiplier Bootstrap}

	\author{Mingjuan Zhang\thanks{ School of Statistics and Mathematics,
Shanghai Lixin University of Accounting and Finance,
Shanghai, China.},~~Yong He\thanks{ School of Statistics, Shandong University of Finance and Economics, Jinan, China; Email:{\tt heyong@sdufe.edu.cn}.}, ~~Cheng Zhou\thanks{ School of Management, Fudan University, Shanghai, China.}, ~~ Xinsheng Zhang\thanks{ School of Management, Fudan University, Shanghai, China.}}	
	\date{}	
	\maketitle
Precision matrix, which is the inverse of covariance matrix, plays an important role in statistics, as it captures the partial correlation between variables. 	Testing the equality of two precision matrices in high dimensional setting is a very challenging  but meaningful problem, especially in the differential network modelling. To our best knowledge, existing test is only powerful for sparse alternative patterns where two precision matrices differ in a small number of elements.
In this paper we propose a data-adaptive test which is powerful against either dense or sparse alternatives. Multiplier bootstrap approach is utilized to approximate the limiting distribution of the test statistic.  Theoretical properties including asymptotic size and power of the test are investigated. Simulation study verifies that the
data-adaptive test performs well under various alternative scenarios. The practical usefulness of the test is illustrated by
 applying it to a gene expression data set associated with lung cancer.

\vspace{2em}


\section{Introduction}
In recent years, Gaussian graphical model has been an important tool to capture the conditional dependency structure among variables. The edges of the Gaussian graphical network are characterized by the inverse covariances for each pair of nodes. To be more specific, for Gaussian graphical model, the joint distribution of $p$ random variables $(X_1,\ldots,X_p)^\top$ is assumed to be multivariate Gaussian $N(\zero,\bOmega^{-1})$, where $\bOmega$ is the inverse of the covariance matrix and is called precision matrix. It is known that for Gaussian graphical model, the conditional dependency structure is completely encoded in the precision matrix, i.e., for each pair of nodes $X_a$ and $X_b$, they are conditionally independent given all other variables if and only if the $(a,b)$-th entry of $\bOmega$ is equal to zero. A growing number of literature has focused on the support recovery and link strength estimation of Gaussian graphical model in high-dimensional setting, see, for example, \cite{meinshausen2006high,yuan2007model,friedman2008sparse,yuan2010high,
	cai2011constrained,cai2012estimating,Liu2013Gaussian}, among many others. For more detailed discussions and comparisons of these methods, we refer to \cite{Ren2015Asymptotic} and \cite{Fan2016Innovated}. The works mentioned above focus on analyzing one particular Gaussian graph. However, in some cases,  it is of greater interest to investigate how the network of connected node pairs change from one state to another. For example, in genomic studies, it is more meaningful to investigate  how the network of connected gene pairs change from different experimental condition,
which provides deeper insights on an underlying biological process, e.g., identification of pathways that correspond to the condition change. Indeed, differential networking modeling has drawn much attention as an important tool to analyze a set of changes in graph structure. The differential network is typically modeled as the difference of two precision matrices and this type of model has been used by \cite{Li2007Finding,Ideker2012Differential, Danaher2014The,zhao2014direct,Xia2015Testing,Tian2016Identifying}. To investigate the differential network, in the first step, we need to identify whether there exists any network change, which is equivalent to test the equality of two precision matrices:,
\begin{equation}\label{hyp:omega2test0}
\Hb_0:\bOmega_1=\bOmega_2.
\end{equation}
{Although the equality of two precision matrices is equivalent to the equality of two covariance matrices from mathematical view, the test problem could be very different due to the fundamental difference between conditional
	and unconditional dependencies.}  Literatures on testing equality of two covariance matrices in high-dimensional setting mainly falls into two categories, sum-of-square type testing and maximum type testing . The sum-of-square type testing are particular powerful under dense alternative where the two covariance matrices differ in a large
number of entries \citep{Schott2007A,Srivastava2010Testing,chen2012two} while the maximum type testing are particular powerful under sparse alternative  where the two covariance matrices differ only in a small
number of entries \citep{Tony2013Two}.  Literature \cite{Zhou2018Unified} proposed a unified framework for developing tests based on U-statistics, which includes testing the equality of two covariance matrices as a special case.  The tests are powerful against a large
variety of alternative scenarios. This research area is very active, and as a result, this list of references is illustrative rather than comprehensive. In contrast, literatures on testing equality of two precision matrices rarely exists. Literature \cite{Xia2015Testing} proposed a maximum-type testing which is powerful against alternative where $\bDelta=\bOmega_1-\bOmega_2$ is sparse.
As far as we know, this is the unique existing work on testing the equality of two precision matrices in the high-dimensional setting. In other word, a powerful testing for hypothesis (\ref{hyp:omega2test0}) under dense alternative still don't exist, which urges us to consider such a problem.

In this paper, we propose a  testing procedure  for hypothesis (\ref{hyp:omega2test0}) which is  powerful against  a large variety of
alternative scenarios in high dimensions. Both theoretical results and numerical simulation show the advantage of proposed test against existing methods.  The rest of the paper is organized as follows.  In section \ref{sec:background},  we introduce some notations and briefly review the test statistic proposed by \cite{Xia2015Testing}.
In Section \ref{sec:methodology} we present our test statistic and the multiplier bootstrap procedure to obtain the critical value or $p$-value of the test.  Section \ref{sec::theoretical} gives the theoretical analysis of the test. In Section \ref{sec:experiments}, we conduct thorough numerical simulation to investigate the empirical performance of the test. A real gene expression data set is analyzed to illustrate the usefulness of the test. At last we discuss possible future directions in the last section.

\section{Background}\label{sec:background}
\subsection{Notation}
For a vector $\vb = (v_1,\dots,v_d)^\top\in \reals^d$, let $\|{\vb}\|_p = \big(\sum_{j=1}^d |v_j|^p\big )^{1/p}$ as the  $L_p$-norm. As $p=\infty$, we set $\|\vb\|_{\infty}=\max_{1\le j\le d}|v_j|$. As $p=0$, we set $\|\vb\|_0=\sum_{j=1}^d I\{v_j\neq0\}$.    We use $v^{(1)}, v^{(2)},\ldots, v^{(d)}$ to denote the order statistics of the absolute value of $\vb$'s entries with $v^{(1)}\le v^{(2)}\le \ldots\le v^{(d)}$. Apparently, we have $v^{(j)}\ge 0$ for $j=1,\ldots,d$. We define the $(s_0,p)$-norm of $\vb$ as $\|{\vb}\|_{(s_0,p)}=\big(\sum_{j=d-s_0+1}^d(v^{(j)})^p\big )^{1/p}$.  As $p=\infty$, we set $\|{\vb}\|_{(s_0,p)}=\|\vb\|_{\infty}=v^{(d)}$ for any $s_0$. We denote
$\mathbb{S}^{d-1}:=\{\vb\in\reals^d:\|\vb\|_2=1\}$ as the spherical surface in $\reals^d$. For any vector $\bmu_m \in \mathbb{R}^d$, let $\bmu_{m,-i}$ denote the $(d-1)\times 1$ vector by removing the $i$-th entry from $\bmu_m$. For a data matrix $\Ub=(\bU_1,\ldots,\bU_n)^\top\in \mathbb{R}^{n\times d}$, let $\Ub_{\cdot,-i}=(\bU_{1,-i},\ldots,\bU_{n,-i})^\top$ with dimension $(n\times(d-1))$, $\bar{\bU}_{\cdot,-i}=n^{-1}\sum_{k=1}^n\bU_{k,-i}$ with dimension $(d-1)\times 1$, $\bU_{(i)}=(U_{1,i},\ldots,U_{n,i})^\top$ with dimension $n\times 1$, $\bar{\bU}_{(i)}=(\bar{U}_i,\ldots,\bar{U}_i)^\top$ with dimension $n\times 1$ where $\bar{U}_i=n^{-1}\sum_{k=1}^nU_{k,i}$ and $\bar{\bU}_{(\cdot,-i)}=(\bar{\bU}_{\cdot,-i},\ldots,\bar{\bU}_{\cdot,-i})^\top$ with dimension $n\times (d-1)$. For tuning parameter $\lambda$, let $\lambda_{n_m,i,m}$ represent the $i$-th tuning parameter for binary trait $m$, which depends on the sample size $n_m$.

For a matrix $\bA=[a_{i,j}]\in \mathbb{R}^{d\times d}$, we denote the matrix $\ell_1$ norm, the matrix element-wise infinity norm and  the matrix element-wise $\ell_1$-norm by $\|\bA\|=\mathrm{max}_{1\leq j\leq d}\sum_{i=1}^d|a_{i,j}|$, $|\bA|_\infty=\max_{i,j}|a_{i,j}|$ and $|\bA|_1=\sum_{i=1}^d\sum_{j=1}^d|a_{i,j}|$ respectively. $\bA_{i,-j}$ denote the $i$-th row of $\bA$ with its $j$-th entry removed and  $\bA_{-i,j}$ denote the $j$-th column of $\bA$ with its $i$-th entry removed. $\bA_{-i,-j}$ denotes a $(d-1)\times(d-1)$ matrix obtained by removing the $i$-th row and $j$-th column of $\bA$. We say $\bA$ is $k$-sparse if each row/column has at most $k$ nonzero entries.
For a symmetric matrix $\bA\in \mathbb{R}^{d\times d}$, we use $\lambda_{\mathrm{min}}(\bA)$ and $\lambda_{\max}(\bA)$ to denote the smallest and largest eigenvalues of $\bA$ respectively. Besides, we define a $d(d-1)/2$-dimension vector
\[
\begin{array}{ll}
{\rm trivec}(\bA)&=(a_{21},  \ldots,a_{d1},a_{32}, \ldots, a_{3d},\ldots, a_{(d-1)d})^\top
\end{array}
\]
which is  obtained by concatenating the lower triangular part of $\bA$ column by column. We use $a_{i_sj_s}$ to denote the $s$-th entry of ${\rm trivec}(\bA)$.

For two sequences of real numbers $\{a_n\}$ and $\{b_n\}$, we write $a_n=O(b_n)$ if there exists a constant $C$ such that $|a_n|\leq C|b_n|$ holds for all $n$,  write $a_n=\bo(b_n)$ if $\lim_{n\rightarrow \infty} a_n/b_n=0$, and write $a_n \asymp b_n$ if there exist  constants $c$ and $C$ such that $c\leq a_n/b_n \leq C$ for all $n$.
For a sequence of random variables $\{\xi_1,\xi_2,\ldots\}$, we use $\lim_{n\rightarrow\infty}\xi_n=\xi$ to denote that the sequence $\{\xi_n\}$ converges in probability towards $\xi$ as $n\rightarrow\infty$. For  simplicity,  we also use $\xi_n=o_p(1)$ to denote $\lim_{n\rightarrow\infty}\xi_n=0$. For random variables $\xi$ and $\eta$, we use ${\rm Cov}(\xi,\eta)$ and ${\rm Corr}(\xi,\eta)$ to denote the covariance and correlation coefficients between $\xi$ and $\eta$. Let ${\rm Var}(\xi)$ be the variance of random variable $\xi$.
For a set $\mathcal{H}$, denote by $\#\{\mathcal{H}\}$ the cardinality of $\mathcal{H}$.

\subsection{General Setup}
Let $\bX=(X_1,\ldots,X_d)^\top$ and $\bY=(Y_1,\ldots,Y_d)^\top$ be two $d$-dimensional  random vectors
independent of each other. $\bX_1,\ldots,\bX_{n_1}$ are independent and identically distributed (i.i.d.)
random samples from $\bX\thicksim N(\bu_1,\bSigma_1)$ with $\bX_k=(X_{k,1},X_{k,2},\ldots,X_{k,d})^\top$. Similarly,  $\bY_1,\ldots, \bY_{n_2}$ are i.i.d. random
samples from $\bY\thicksim (\bu_2,\bSigma_2)$ with $\bY_k=(Y_{k,1},Y_{k,2},\ldots,Y_{k,d})^\top$.  Let $\Xb=(\bX_1,\ldots,\bX_{n_1})^\top$ and $\Yb=(\bY_1,\ldots,\bY_{n_1})^\top$ denote the data matrices. Let $\bSigma_m=(\sigma_{i,j,m})$ and $\bOmega_m=(\omega_{i,j,m})=\bSigma_m^{-1}$ for $m=1,2$. Let $\bbeta_{i,1}=(\beta_{1,i,1},\ldots,\beta_{d-1,i,1})^\top$ denote the regression coefficients of $X_{k,i}$ regressed on the rest of the entries of $\bX_k$ and let $\bbeta_{i,2}=(\beta_{1,i,2},\ldots,\beta_{d-1,i,2})^\top$ denote the regression coefficients of $Y_{k,i}$ regressed on the rest of the entries of $\bY_k$.

In the Gaussian  setting, the precision matrix can be described in terms of regression models. Specifically:
\begin{equation}\label{PrecisonExpression1}
\begin{aligned}
X_{k,i}&=\alpha_{i,1}+\bX_{k,-i}^\top\bbeta_{i,1}+\epsilon_{k,i,1}, \\
Y_{k,i}&=\alpha_{i,2}+\bY_{k,-i}^\top\bbeta_{i,2}+\epsilon_{k,i,2}, 
\end{aligned}
\end{equation}
where the error terms $\epsilon_{k,i,m}$ follow normal distribution with mean zero and variance $$\{\sigma_{i,i,m}-\bSigma_{i,-i,m}(\bSigma_{-i,-i,m})^{-1}\bSigma_{-i,i,m}\}.$$ and $\epsilon_{k,i,1}$, $\epsilon_{k,i,2}$ are independent of $\bX_{k,-i}$ and $\bY_{k,-i}$ respectively.

Besides, we have $\alpha_{i,m}=\mu_{i,m}-\bSigma_{i,-i,m}\bSigma_{-i,-i,m}^{-1}\bmu_{-i,m}$. The regression coefficient vectors $\bbeta_{i,m}$ and error terms $\epsilon_{k,i,m}$ satisfy
\begin{equation*}
\begin{aligned}
\bbeta_{i,m}=-\omega_{i,i,m}^{-1}\bOmega_{-i,i,m}, \qquad \qquad
r_{i,j,m}=\Cov\big(\epsilon_{k,i,m},\epsilon_{k,j,m}\big)=\frac{\omega_{i,j,m}}{\omega_{i,i,m}\omega_{j,j,m}}.
\end{aligned}
\end{equation*}
We aim to test the null hypothesis:
\[
\Hb_0:\bOmega_1=\bOmega_2 \ \  \mathrm{or} \ \ \mathrm{equivalentlly } \ \  \bDelta=\bOmega_1-\bOmega_2=0.
\]
Let $\hbbeta_{i,m}=(\widehat{\beta}_{1,i,m},\ldots,\widehat{\beta}_{d-1,i,m})^\top$ be estimators of $\bbeta_{i,m}$ by Lasso or Dantzig selector satisfying
\begin{equation}\label{betaestimatebound}
\begin{aligned}
\bmax_{1\leq i\leq d}\big\|\hbbeta_{i,m}-\bbeta_{i,m}\big\|_1=\bop\big\{(\log d)^{-1}\big\},
\qquad \qquad
\mathop{\mathrm{\max}}_{1\leq i\leq d}\big\|\hbbeta_{i,m}-\bbeta_{i,m}\big\|_2=\bop\big\{(n_m\log d)^{-1/4}\big\}.
\end{aligned}
\end{equation}
Under the sparsity conditions $\max_{1\le i\le d}|\beta_{i}|_{0}=\bo\big(n^{1/2}/(\log d)^{3/2}\big)$,  together with with Assumption \textbf{(B)} in Section \ref{subsec::assumption},  both the Lasso and Dantzig selector estimators satisfy the condition in \eqref{betaestimatebound} according to the Proposition 4.1 in \cite{Liu2013Gaussian}.

With the $\hbbeta_{i,m}$, define the residuals by
\begin{equation}\label{equ:residual}
\begin{aligned}
\hepsilon_{k,i,1}&=X_{k,i}-\bar{X}_i-(\bX_{k,-i}-\bar{\bX}_{\cdot,-i})^\top\hbbeta_{i,1}, \\ \hepsilon_{k,i,2}&=Y_{k,i}-\bar{Y}_i-(\bY_{k,-i}-\bar{\bY}_{\cdot,-i})^\top\hbbeta_{i,2}.
\end{aligned}
\end{equation}
Let $\widetilde{r}_{i,j,m}=(1/{n_m})\sum_{k=1}^{n_m}\hepsilon_{k,i,m}\hepsilon_{k,j,m}$ be the empirical covariance between $\{\hepsilon_{k,i,m}:k=1,\ldots,n_m\}$ and $\{\hepsilon_{k,j,m}:k=1,\ldots,n_m\}$. Similarly, let $\widetilde{R}_{i,j,m}=(1/{n_m})\sum_{k=1}^{n_m}(\epsilon_{k,i,m}-\bar{\epsilon}_{i,m})(\epsilon_{k,j,m}-\bar{\epsilon}_{j,m})$ be the empirical covariance between $\{\epsilon_{k,i,m}:k=1,\ldots,n_m\}$ and $\{\epsilon_{k,j,m}:k=1,\ldots,n_m\}$.
Lemma 2 in \cite{Xia2015Testing} shows that
\begin{equation}\label{rhatRt}
\begin{aligned}
\widetilde{r}_{i,j,m}=\widetilde{R}_{i,j,m}-\widetilde{r}_{i,i,m}
(\widehat{\beta}_{i,j,m}-\beta_{i,j,m})-\widetilde{r}_{j,j,m}(\widehat{\beta}_{j-1,i,m}-\beta_{j-1,i,m})+\bop\big\{
(n_m\log d)^{-1/2}\big\}.
\end{aligned}
\end{equation}
For $1\leq i<j\leq d$, it can be shown that
$$\beta_{i,j,m}=-\omega_{i,j,m}/\omega_{j,j,m}, \ \ \beta_{j-1,i,m}=-\omega_{i,j,m}/\omega_{i,i,m}.$$
A bias-corrected estimator of  $r_{i,j,m}(1\leq i<j\leq d)$ is initially proposed by \cite{Liu2013Gaussian}:
\begin{equation}\label{rijmestimatebyXia}
\widehat{r}_{i,j,m}\!=\!\!-\!\big(\widetilde{r}_{i,j,m}\!+\!\widetilde{r}_{i,i,m}\widehat{\beta}_{i,j,m}\!+\!\widetilde{r}_{j,j,m}\widehat{\beta}_{j\!-\!1,i,m}\big), 
\end{equation}
For $i=j$, the Lemma 2 of \cite{Xia2015Testing} showed that
\begin{equation}\label{tilderriibound}
\mathop{\mathrm{\max}}_{1\leq i\leq d}\big|\widetilde{r}_{i,i,m}-r_{i,i,m}\big|=\bOp\big\{(\log d/n_m)^{1/2}\big\},
\end{equation}
which implies that $\widehat{r}_{i,i,m}=\widetilde{r}_{i,i,m}$ is a nearly unbiased estimator of $r_{i,i,m}$.
Thus one can naturally estimate $\omega_{i,j,m}$ by
\begin{equation}\label{def:Tijm}
T_{i,j,m}=\frac{\widehat{r}_{i,j,m}}{\widehat{r}_{i,i,m}\widehat{r}_{j,j,m}}
, 1\leq i\leq j\leq d,
\end{equation}
and test $\Hb_0:\bDelta=0$ based on the estimators $\mathcal{T}=\{T_{i,j,1}-T_{i,j,2},1\leq i\leq j\leq d\}$.

Considering the heteroscedasticity  of estimators in $\mathcal{T}$, Literature \cite{Xia2015Testing}  proposed the following test statistic for the null hypothesis $\Hb_0$:
\[
M_n=\mathop{\mathrm{\max}}_{1\leq i \le j \leq d}W_{i,j}^2=\mathop{\mathrm{\max}}_{1\leq i\le j\leq d}\frac{(T_{i,j,1}-T_{i,j,2})^2}{\widehat{\theta}_{i,j,1}+\widehat{\theta}_{i,j,2}},
\]
where
\begin{equation}\label{Wij}
\begin{aligned}
W_{i,j}=\frac{T_{i,j,1}-T_{i,j,2}}{(\widehat{\theta}_{i,j,1}+\widehat{\theta}_{i,j,2})^{1/2}}, \qquad \qquad  \widehat{\theta}_{i,j,m}=\text{Var}(T_{i,j,m})=
\frac{1+\widehat{\beta}_{i,j,m}^2\widehat{r}_{i,i,m}/\widehat{r}_{j,j,m}}{n_m\widehat{r}_{i,i,m}\widehat{r}_{j,j,m}}.
\end{aligned}
\end{equation}

Literature \cite{Xia2015Testing} obtained the asymptotic null distribution of $M_n$ under suitable conditions, which is type I extreme value distribution. However, this limiting distribution of maximum-type statistic based approach has
two fatal limitations.   Firstly, the convergence rate of extreme-value statistics is notoriously slow
and the process of getting the limiting distribution ignores the correlation between coordinates. Secondly, the maximum-type statistic is particularly powerful against large and sparse signal alternatives, however, it is powerless against small and dense signal alternatives.

In this paper we develop new tests for hypothesis (\ref{hyp:omega2test0}), which
are adaptive to a large variety of alternative scenarios in high dimensions. We utilize the multiplier bootstrap method to approximate the asymptotic distribution of the proposed test statistics and thus overcomes the limitation of the extreme-value-type statistic $M_n$.

\section{Methodology}\label{sec:methodology}
As the extrem-value-type statistic is only powerful against the sparse large alternatives, we aim to provide a data-driven adaptive test for the hypothesis (\ref{hyp:omega2test0}) in this section. In Section \ref{sec:individual test}, a family of tests based on $(s_0,p)$-norm are proposed. The $(s_0,p)$-norm  was first introduced in \cite{Zhou2018Unified}. The tests based on different $p$ have different powers under different alternative scenarios. For example, $(s_0,\infty)$-norm based test are  sensitive to large  perturbations on a small number of entries of $\bOmega_1-\bOmega_2$. Moreover, $(s_0,2)$-norm are sensitive to small perturbations on  a large number of entries of $\bOmega_1-\bOmega_2$. By combining a family of $(s_0,p)$-norm based tests with various $p$, we present our  adaptive test in Section \ref{sec:adaptive test}

\subsection{The $(s_0,p)$-norm based test statistics}\label{sec:individual test}
In this section, we provide some ($s_0, p$)-norm based tests. Recall that we have defined the $W_{i,j}$ in (\ref{Wij}). Based on the statistics in $\mathcal{W}=\{W_{i,j},1\leq i\leq j\leq d\}$, define $\Wb=(W_{i,j})_{d\times d}$. we then propose our test statistic based on $(s_0,p)$-norm of the vector $\mathrm{trivec}(\Wb)$. Specifically, we propose the $(s_0,p)$-norm based test statistic is
\begin{equation}\label{teststatistic}
N_{(s_0,p)}=\big\|\mathrm{trivec}(\Wb)\big\|_{(s_0,p)}.
\end{equation}
With the proposed test statistic, we still need to obtain the critical value or $P$-value to test (\ref{hyp:omega2test0}). To this end, we develop a multiplier bootstrap method to approximate the limiting distribution of the test statistic $N_{(s_0,p)}$.

In the high dimensional setting, \cite{chernozukov2014central} introduced the  multiplier bootstrap method for the sum of independent random vectors. In detail, let $\bZ_1, \ldots,\bZ_{n}$ be independent zero mean random vectors in $\reals^d$ with
$\bZ_k=(Z_{k1},\ldots,Z_{kd})^\top$. The bootstrap sample for the sample mean
$n^{-1}\sum_{k=1}^n\bZ_k$ then becomes $n^{-1}\sum_{k=1}^n\varepsilon_k\bZ_k$, where $\varepsilon_1, \varepsilon_2,\ldots,\varepsilon_n$ are independent standard normal random variables. Inspired by the multiplier bootstrap method in \citep{chernozukov2014central}, we propose a specific multiplier bootstrap procedure for the problem here.  In detail, we generate independent samples $\eta_{1,1}^b,\ldots,\eta_{1,n_1}^b$ and
$\eta_{2,1}^b,\ldots,\eta_{2,n_2}^b$ from $\eta\sim N(0,1)$ for $b=1,\ldots,B$.  Similarly, we set the $b$-th multiplier bootstrap sample for $\widetilde{r}_{i,j,m}, 1\leq i\leq j\leq d$ as
\begin{equation}\label{def:originalbootsr}
\widetilde{r}_{i,j,m}^b=\frac{1}{n_m}\sum_{k=1}^{n_m}\eta_{m,k}^b(\hepsilon_{k,i,m}\hepsilon_{k,j,m}-\widetilde{r}_{i,j,m}).
\end{equation}
Considering the definitions of $\widehat{r}_{i,j,m}$ in \eqref{rijmestimatebyXia}, we set its $b$-th bootstrap sample as
\[
\widehat{r}_{i,j,m}^b=-\big(\widetilde{r}_{i,j,m}^b+\widetilde{r}_{i,i,m}^b\widehat{\beta}_{i,j,m}+
\widetilde{r}_{j,j,m}^b\widehat{\beta}_{j-1,i,m}\big)
\]
for $1\leq i<j\leq d$ and  $\widehat{r}_{i,i,m}^b=\widetilde{r}_{i,j,m}^b$.

Further, by the definitions of $T_{i,j,m}$ and $W_{i,j}$ in \eqref{def:Tijm} and \eqref{Wij} respectively, we then get the $b$-th bootstrap sample of $T_{i,j,m}$ and $W_{i,j}$ as
\begin{equation}\label{Wbootsineqj}
\begin{aligned}
T_{i,j,m}^b=\frac{\widehat{r}_{i,j,m}^b}{\widehat{r}_{i,i,m}\widehat{r}_{j,j,m}}, 1\leq i\leq j\leq d,  \qquad \qquad
W_{i,j}^b=\frac{T_{i,j,1}^b-T_{i,j,2}^b}{(\widehat{\theta}_{i,j,1}+\widehat{\theta}_{i,j,2})^{1/2}}, 1\leq i\leq j\leq d.
\end{aligned}
\end{equation}
With $W_{i,j}^b$, we set $\Wb^b=(W_{i,j}^b)_{d\times d}$ and finally obtain the bootstrap samples of $N_{(s_0,p)}$ as
\begin{equation}\label{test-statistic-bootstrap-sample}
N_{(s_0,p)}^b=\|\mathrm{trivec}(\Wb^b)\|_{(s_0,p)}, \hspace{2em} b=1\ldots B.
\end{equation}
Given the significance level $\alpha$,
we use $t^{N}_{\alpha,(s_0,p)}$  to denote the oracle  critical values of $N_{(s_0,p)}$ . Given the bootstrap samples, we then estimate  $t^{N}_{\alpha,(s_0,p)}$  by
\begin{equation}
\hat{t}^{N}_{\alpha,(s_0,p)}=\inf\Big\{t\in \reals: \frac{1}{B}\sum_{b=1}^B\ind \{N^b_{(s_0,p)}\le t\}> 1-\alpha\Big\}.
\end{equation}
Therefore, we obtain the $(s_0,p)$-norm based tests for (\ref{hyp:omega2test0}) as
\begin{equation}
T^{N}_{(s_0,p)}=\ind\big\{N_{(s_0,p)}\ge \hat{t}^N_{\alpha,(s_0,p)}\big\}.
\end{equation}
We reject $\Hb_0$ of (\ref{hyp:omega2test0}) if and only if $T^N_{(s_0,p)}=1$.
Accordingly, we estimate  $N_{(s_0,p)}$'s oracle  $P$-values $P^N_{(s_0,p)}$ by
\begin{equation}\label{def:PhatN}
\hat{P}^N_{(s_0,p)}=\frac{\sum_{b=1}^B\ind\{N^b_{(s_0,p)}> N_{(s_0,p)}\}}{B+1}.
\end{equation}
Therefore, given a significance level $\alpha$,  we reject $\Hb_{0}$ of (\ref{hyp:omega2test0}) if and only if $\hat{P}^N_{(s_0,p)}\le
\alpha$.

\begin{algorithm}[ht]
	\caption{A bootstrap procedure to obtain $N_{\rm ad}$}\label{alg:first}
	\raggedright
	{\bf Input:} $\mathcal{X}$.\\
	{\bf Output:}  $N^1_{(s_0,p)},\ldots,N^B_{(s_0,p)}$ with $p\in \mathcal{P}$, and $N_{\rm ad}$.\\
	\begin{algorithmic}[1]
		\Procedure{}{}
		\State $N_{(s_0,p)} \!=\! \|{\mathrm{trivec}(\Wb)}\|_{(s_0,p)} \text{ with } {\Wb} \!=\! (W_{i,j})_{d\times d}^\top$ and $W_{i,j} \!=\! {(T_{i,j,1} \!-\! T_{i,j,2})}\big/{(\widehat{\theta}_{i,j,1} \!+\! \widehat{\theta}_{i,j,2})^{1/2}}$.
		\For{ $b\leftarrow 1$ {\bf to} $B$}
		\State  Sample independent standard normal random variables $\{\eta^b_{1,1},\ldots,\eta^b_{1,n_m}\}$ for $m=1,2$.
		\State  For $1\leq i\leq j\leq d$, set $\widetilde{r}_{i,j,m}^b=({1}/{n_m})\sum_{k=1}^{n_m}\eta_{m,k}^b\big(\hepsilon_{k,i,m}\hepsilon_{k,j,m}-\widetilde{r}_{i,j,m}\big)$.
		
		\State Set $\widehat{r}_{i,i,m}^b=\widetilde{r}_{i,i,m}^b$, and set $\widehat{r}_{i,j,m}^b\!=\!\!-\!\big(\widetilde{r}_{i,j,m}^b\!+\!\widetilde{r}_{i,i,m}^b\widehat{\beta}_{i,j,m}\!+\!\widetilde{r}_{j,j,m}^b\widehat{\beta}_{j-1,i,m}\big)$ for $1\leq i<j\leq d$.

		\State  Set $T_{i,j,m}^b\!=\!{\widehat{r}_{i,j,m}^b}\big/({\widehat{r}_{i,i,m}\widehat{r}_{j,j,m}})$, $1\leq i\leq j\leq d$.
		\State Set $W_{i,j}^b=({T_{i,j,1}^b-T_{i,j,2}^b})\big/{(\widehat{\theta}_{i,j,1}+\widehat{\theta}_{i,j,2})^{1/2}}$, $1\leq i\le j\leq d$ and $\Wb^b=(W_{i,j}^b)_{d\times d}$.
		
		\For{$p$ {\bf in} $\mathcal{P}$}
		\State $N^b_{(s_0,p)}=\|{\mathrm{trivec}(\Wb}^b)\|_{(s_0,p)}$ with ${\Wb}^b=(W_{i,j}^b)_{d\times d}$.
		\EndFor
		\EndFor
		\State $\hat{P}^N_{(s_0,p)}={\sum_{b=1}^B\ind\{N^b_{(s_0,p)}> N_{(s_0,p)}\}}/{(B+1)}$ for $p\in \mathcal{P}$.
		\State $N_{\rm ad}=\min_{p\in\mathcal{P}}\hat{P}^N_{(s_0,p)}$.
		\EndProcedure
	\end{algorithmic}
\end{algorithm}

\subsection{Data adaptive combined test}\label{sec:adaptive test}
After providing the $(s_0,p)$-norm based tests for each individual $p$, we propose a data-driven adaptive test by combining a group of the $(s_0,p)$-norm based tests in this section.

Set $\mathcal{P}=\{p_1,p_2,\cdots\}$ as a finite set of positive numbers, and set the size of $\cP$ as a finite fixed constant. Then we combine the $(s_0,p)$-norm based test with $p \in \mathcal{P}$ by taking the minimum $P$-value of these tests. Specifically, we set the data-adaptive test statistic $N_{\rm ad}$ as
\begin{equation}\label{def:Nad}
N_{\rm ad}=\min_{p\in\mathcal{P}}\hat{P}^N_{(s_0,p)}.
\end{equation}
The detail process of getting $N_{\rm ad}$ is in Algorithm \ref{alg:first}.
The set $\{\mathcal{P}\}$ can be chosen by users with  prior information about the alternative patterns. If one knows the alternative pattern, then he/she can choose the set $\mathcal{P}$ accordingly to improve the power performance of the data adaptive test. For example,  let $\mathcal{P}$ consists of large values of $p$ with prior information that the  alternative pattern is sparse. If one knows nothing about the alternative pattern, then  a balanced set $\mathcal{P}$ containing both large and small $p$ is recommended. For example, one may choose the set $\mathcal{P}$ to be $\{1,2,3,4,5,\infty\}$.

For the data adaptive test, we  need to get the $P$-value. It's difficult to get the limiting distribution for the $(s_0,p)$-norm based statistics, not to mention  for the data adaptive test statistic. Hence, the intuitive way is to do a double loop bootstrap procedure to get the empirical distribution for our data adaptive test. But this way is too costly for computation. As is shown by Algorithm \ref{alg:first}, in addition to the data adaptive statistic $N_{\rm ad}$, we also obtain the bootstrap samples for  $N_{(s_0,p)}$, i.e, $\big\{N^1_{(s_0,p)},\ldots,N^B_{(s_0,p)}\big\}$. Therefore, we can recycle the bootstrap samples to accelerate our computation speed. Specifically, for $b=1,\ldots,B$ and $p\in\mathcal{P}$, we set
\[
\hat{P}^{b,N}_{(s_0,p)}=\frac{\sum_{b_1\ne b} \ind\{N^{b_1}_{(s_0,p)}> N^{b}_{(s_0,p)}\}}{B}.
\]
We use $N^b_{\rm ad}=\min_{p\in\mathcal{P}
}\hat{P}^{b,N}_{(s_0,p)}$  as the bootstrap sample for $N_{\rm ad}$. We then
estimate the oracle $P$-value of $N_{\rm ad}$  by
\begin{equation}\label{def:hatPWad}
\hat{P}_{\rm ad}^N = \frac{\big(\sum_{b=1}^{B}\ind\{N^b_{\rm ad}\le N_{\rm ad}\}\big)+1}{B+1}.
\end{equation}
For more details, see Algorithm \ref{alg:adjust}. The samples  $N^1_{\rm ad},\ldots,N^B_{\rm ad}$ are nonindependent. But as $n,B\rightarrow\infty$, they are asymptotically independent. Hence, it dosen't affect the consistency of $\hat{P}^N_{\rm ad}$.
After getting the estimated $P$-values of the data-adaptive tests  $N_{\rm ad}$, given the significance level $\alpha$, we set
\begin{equation}\label{def:TWNAd}
T^N_{\rm ad}=\ind\{\hat{P}^N_{\rm ad}\le\alpha\}.
\end{equation}
Therefore,  we reject $\Hb_0$ of (\ref{hyp:omega2test0}) if and only if $T^N_{\rm ad}=1$.

\begin{algorithm}[ht]
	\caption{A low-cost  bootstrap procedure}\label{alg:adjust}
	\raggedright
	{\bf Input:} $\mathcal{X}$ and $N^1_{(s_0,p)},\ldots,N^B_{(s_0,p)}$ for $p\in \mathcal{P}$.\\
	\raggedright
	{\bf Output:} $N_{\rm ad}^1,\ldots, N_{\rm ad}^B$.
	\begin{algorithmic}[1]
		\Procedure{}{}
		\For{$b \leftarrow 1$ {\bf to} $B$}
		\For{ $p$ {\bf in} $\mathcal{P}$}
		\State $\hat{P}^{b,N}_{(s_0,p)}={\sum_{b_1\ne b} \ind\{N^{b_1}_{(s_0,p)}> N^{b}_{(s_0,p)}\}}/{B}$.
		\EndFor
		\State $N^b_{\rm ad}=\min_{p\in\mathcal{P}}\hat{P}^{b,N}_{(s_0,p)}$.
		\EndFor
		\EndProcedure
	\end{algorithmic}
\end{algorithm}

\section{Theoretical properties}\label{sec::theoretical}
In this section, we investigate the theoretical properties of our proposed test. Firstly, some  assumptions are introduced in Section  \ref{subsec::assumption}.  In Section  \ref{subsec:theoretical}, we verify the validity of multiplier bootstrap which is used in Section  \ref{sec:methodology} and then analyze the theoretical properties of the proposed test.

\subsection{Assumptions}\label{subsec::assumption}
In this section, we introduce some assumptions that are commonly used in high-dimensional analysis.

\vspace{0.8em}
\textbf{(A)} Set $n =\mathrm{max}(n_1,n_2)$, there exists some $0<\delta<1/7$ such that $s_{0}^{2}\log(d)=\bo(n^{\delta})$ hold, where $n_1\asymp n_2\asymp n$.
\vspace{0.8em}

Assumption \textbf{A} allows $s_0$ and $d$ go to infinity as long as $s_{0}^{2}\log(d)=\bo(n^{\delta})$ hold. By using the multiplier bootstrap to get the critical values for our tests, we need some more assumptions compared with \cite{Xia2015Testing}. Other than the Assumption \textbf{(A)}, we also introduce a more strong Assumption \textbf{(A)$'$} to state the scaling of $s_0$, $d$ and $n$. Before stating the next assumption, we need some additional notations. Let $U_{i,j,m}=\frac{1}{n_{m}} \sum_{k=1}^{n_m} \Big\{ \epsilon_{k,i,m} \epsilon_{k,j,m} - \mathbb{E}\big(\epsilon_{k,i,m}\epsilon_{k,j,m}\big) \Big\}$ and define $\tilde{U}_{i,j,m}=(r_{i,j,m}-U_{i,j,m})/r_{i,i,m}r_{j,j,m}$ with $1\le i,j\le d$, $m=1,2$.
Define $\tilde{\Ub}_{m}=\big( \tilde{U}_{i,j,m}\big)$ as a square matrix of order $d$ and denote
the covariance matrix of $\text{trivec}(\tilde{\Ub}_{m})$ as $\bSigma_{m}^{\tilde{U}}=(\sigma^{\tilde{U}}_{s,t,m})_{1\le s,t\le d(d-1)/2}$, where
\[
\sigma^{\tilde{U}}_{s,t,m} = \left\{\
\begin{aligned}
\theta_{i,j,m}=
\frac{1+ \beta_{i,j,m}^{2} r_{i,i,m}/r_{j,j,m}}{n_m r_{i,i,m}r_{j,j,m}},  s=t,\\
\frac{r_{i_{1},i_{2},m} r_{j_{1},j_{2},m}+r_{i_{1 }, j_{2},m}r_{i_{2},j_{1},m}}{n_m r_{i_1,i_1,m} r_{j_1,j_1,m} r_{i_2,i_2,m} r_{j_2,j_2,m}},  s\neq t,
\end{aligned}
\right.
\]
with $1\le i<j\le d$, $1\le i_1<j_1\le d$, $1\le i_2<j_2\le d$, $i_1\neq i_2$, $j_1\neq j_2$, $m=1,2$.

Let $\bG$ be a Gaussian random vector in $\reals^{d(d-1)/2}$ with mean zero and covariance matrix $\mathbf{R}^{\tilde{U}}_{12}$, where $\mathbf{R}^{\tilde{U}}_{12}=(\mathbf{D}^{\tilde{U}}_{12})^{-1/2}\bSigma^{\tilde{U}}_{12}(\mathbf{D}^{\tilde{U}}_{12})^{-1/2}$ with $\bSigma^{\tilde{U}}_{12}=\bSigma^{\tilde{U}}_1/n_1+\bSigma^{\tilde{U}}_2/n_2$ and $\mathbf{D}^{\tilde{U}}_{12}=\mathrm{Diag}(\bSigma^{\tilde{U}}_{12})$. Set the probability density function (PDF) and the  $\alpha$-quantile of $\|\bG\|_{(s_0,p)}$ as $f_{\bG,(s_0,p)}$ and $c_{ (s_0,p)}(\alpha)$ respectively. We then define $h_{T}(z)$ as
\begin{equation*}
\begin{aligned}
h_{T}(z)=\max_{p\in \mathcal{P}}\max_{x\in C_{(s_0,p)}(z)}f^{-1}_{\bG,(s_0,p)}(x) \\
\text{with}\quad C_{(s_0,p)}(z)=[c_{(s_0,p)}(z),c_{(s_0,p)}(1-z)].
\end{aligned}
\end{equation*}
With these new notations, we introduce the following assumption.

\vspace{0.8em}
\textbf{(A)$'$} Define $n =\mathrm{max}(n_1,n_2)$. We assume that $h_{T}^{0.6}(z)s_{0}^{2}\log d=\bo(n^{1/10})$ holds for any $0<z<1$ as $n, d\rightarrow\infty$ and $n_1\asymp n_2\asymp n$.
\vspace{0.8em}

{Assumption {\bf (A)$'$} is more stringent. It is critical to guarantee the uniform convergence of the distribution functions and the corresponding quantile functions of the test statistics $N_{(s_0,p)}$ for any $p \in \mathcal{P}$. }
The next two mild assumptions are often used in high dimensional setting, especially when the inference for covariance matrix and precision matrix are involved.

\vspace{0.8em}
\textbf{(B)} {
	There exist some positive constants $C_0<C_1$,
	such that $\lambda_{\min}(\bOmega_m)\ge C_0$ and $\lambda_{\max}(\bOmega_m) \le C_1$, with $m=1,2$.}
There exists some $\tau>0$ such that $|A_\tau| = \bo(d^{1/16})$ where $A_{\tau}= \{(i,j):|w_{i,j,m}|\ge(\log d)^{-2-\tau},1\le i<j\le d, \text{ for } m=1 \text{ or } 2\}$.
\vspace{0.8em}

\vspace{0.8em}
\textbf{(C)} Let $D_m$ be the diagonal of $\bOmega_{m}$ and let $(\eta_{i,j,m})=D^{-1/2}_{m}\bOmega_{m}D^{-1/2}_{m}$, for $m=1,2$. Assume that $\max_{1\le i\le j\le d}|\eta_{i,j,m}|\le \eta_{m}\le c$, where $0<c<1$ is a constant.

\vspace{0.8em}

{\textbf{(D)} Suppose $\max_{1\le i\le d}s_{i,m}=\big(n^{1/2}/(\log d)^{3/2}\big)$, where $s_{i,m}$ is sparsity for the $i$-th row or column of $\bOmega_{m}$ for $m=1,2$.
}

{Note that $\bbeta_{i,m}=-\omega_{i,i,m}^{-1}\bOmega_{-i,i,m}$, then the sparsity conditions of the Proposition 4.1 in \cite{Liu2013Gaussian} are automatically satisfied under Assumption \textbf{(D)}.}

\subsection{Theoretical analysis}\label{subsec:theoretical}
After introducing some needing assumptions, we analyze the theoretical properties of our test. Due to the complicated structure of our test statistics, we use the multiplier bootstrap to get the critical values for our test in Section \ref{sec:methodology}. But this procedure is different from \cite{Zhou2018Unified}. Specifically, other than the testing statistics cannot be rewritten as a sum of independent random variables, there are also some bias correction terms. Hence, we need to justify the validity of this multiplier bootstrap.
\begin{theorem}\label{theorem:bootstrap}
	Suppose Assumptions {\bf (A)}-{\bf(D)} hold. Under the null hypothesis $\Hb_0$ of (\ref{hyp:omega2test0}), we have as $n, d\rightarrow\infty$,
	\begin{equation}\label{equ:bootstrap}
	\mathop{\mathrm{sup}} \limits_{z\in(0,\infty)}\Big|\P(N_{(s_0,p)}\leq z)\!-\!\P(N^{b}_{(s_0,p)} \!\leq\! z|\mathcal{X},\mathcal{Y})\Big|\!=\!\bo(1).
	\end{equation}
\end{theorem}
Under the Gaussian distribution setting, it is easily to check that the sub-exponential distribution assumption and the moment condition in \cite{chernozukov2014central} are satisfied. Hence, under milder conditions, Theorem \ref{theorem:bootstrap} verifies the validity of the multiplier bootstrap method. The proof of Theorem \ref{theorem:bootstrap} is in the Appendix.

{By Theorem \eqref{theorem:bootstrap} hold, it's easy to prove that the size of $T^{N}_{(s_0,p)}$  asymptotically coverges to pre-specified significance level $\alpha$.}
\begin{corollary}
	Suppose Assumptions {\bf (A)}-{\bf(D)} hold. Under the null hypothesis $\Hb_0$ of (\ref{hyp:omega2test0}), we have
	\begin{equation*}
	\mathbb{P}_{\Hb_0}(T^{N}_{(d_0,p)}=1)\rightarrow \alpha,
	\end{equation*}
	as $n, B\rightarrow\infty$.
\end{corollary}

With Theorem \ref{theorem:bootstrap}, we then show the theoretical properties of our data-adaptive test $T^{N}_{\rm ad}$. By the definition of $\hat{P}_{\rm ad}^N$ and $T_{\rm{ad}}^{N}$ in \eqref{def:hatPWad} and \eqref{def:TWNAd}, it can be seen that $T^{N}_{\rm ad}$ relies on the estimated $P$-values of $N_{(s_0,p)}$.
Therefore,  we suppose the more stringent Assumption {\bf (A)$'$} holds, which guarantees the uniform convergence of the distribution
functions and the corresponding quantile functions of the test statistics $N_{(s_0,p)}$ for any $p \in \mathcal{P}$. Under this condition, we show that the empirical size of the data-adaptive test $T^{N}_{\rm ad}$ approximates to the pre-specified level $\alpha$.
\begin{theorem}\label{theorem:sizead}
	Suppose Assumptions \textbf{(A)$'$}, \textbf{(B)}-\textbf{(D)}  hold. Under the null hypothesis $\Hb_0$ of (\ref{hyp:omega2test0}), we have
	\begin{equation*}
	\mathbb{P}_{\Hb_0}(T^{N}_{\rm ad}=1)\rightarrow \alpha,
	\end{equation*}
	as $n, B\rightarrow\infty$.
\end{theorem}

After analyzing the asymptotic sizes of {the tests in Section \ref{sec:methodology},}
we summarize the asymptotic power properties in the following theorem. To analyze the power performance of $T^{N}_{\rm ad}$, we need to introduce some other notations.  Define $\Wb^{*}=(W^{*}_{i,j})_{d\times d}$ with
\begin{equation*}\label{def:powerAssumption}	
W^{*}_{i,j}=\big|\tilde{U}_{i,j,1}-\tilde{U}_{i,j,2}\big|\Big/\sqrt{\theta_{i,j,1}/n_1+\theta_{i,j,2}/n_2},
\end{equation*}
where $\theta_{i,j,m}$ are the diagonal elements of $\bSigma_{m}^{\tilde{U}}$, $m=1,2$. We then introduce the following theorem to characterize the asymptotic power properties of {$T^{N}_{(s_0,p)}$ and} $T^{N}_{\rm ad}$.

\begin{theorem}\label{thm:powerad}
	Suppose  Assumptions {\bf (B)}-{\bf(D)} hold and assume  $\varepsilon_n=\bo(1)$,  $\varepsilon_n\sqrt{\log d^2}\rightarrow\infty$ as $n,d\rightarrow\infty$.
	
	\textbf{(a)} As $n,d \rightarrow \infty$, there exists some $\delta_{1}>0$ such that $\log(d)=\bo(n^{1/3})$ and $n=\bO(d^{2\delta_{1}})$. Assume $s_0=\bO\big((\log d)^{\delta_{2}}\big)$ for some postive constant $\delta_{2}$. Under the alternative hypothesis $\Hb_1$ of (\ref{hyp:omega2test0}) and with
	\begin{equation*}
	\begin{aligned}
	\big\|{\rm trivec}(\Wb^{*})\big\|_{(s_0,p)}\ge s_0 (1+\varepsilon_n)
	\big(\sqrt{2\log(d(d-1)/2)}+\sqrt{2\log (1/\alpha)}\big),
	\end{aligned}
	\end{equation*}
	hold, we  have $\P_{\Hb_1}\big(T^{N}_{(s_0,p)}=1\big)\rightarrow 1$ as $n, d, B\rightarrow \infty$.
	
	\textbf{(b)} With Assumption {\bf (A)$'$} hold, and under the alternative hypothesis $\Hb_1$ of (\ref{hyp:omega2test0}) and suppose
	\begin{equation*}
	\begin{aligned}
	\big\|{\rm trivec}(\Wb^{*})\big\|_{(s_0,p)}\!\ge\! s_0 (1\!+\!\varepsilon_n)\big(\sqrt{2\log(d(d\!-\!1)/2)}+\sqrt{2\log (\#\{\mathcal{P}\}/\alpha)}\big),
	\end{aligned}
	\end{equation*}
	hold, we  have $\P_{\Hb_1}\big(T^{N}_{\rm ad}=1\big)\rightarrow 1$ as $n, d, B\rightarrow \infty$.
\end{theorem}
By Theorem \ref{thm:powerad}, we show that the asymptotic
{ powers of $T^{N}_{(s_0,p)}$ and $T^{N}_{\rm ad}$ converge to 1 under the minimum signal condition on $\|{\rm trivec}(\Wb^{*})\|_{(s_0,p)}$.}

\section{Experiments}\label{sec:experiments}
\subsection{Simulation study}	\label{subsec:simulation}
In this section, we conduct simulation study to investigate the empirical size and power of the proposed test. To show the adaptivity of our method, we compare it with recently developed method proposed by \cite{Xia2015Testing} under various model settings.
We denote the test proposed by \cite{Xia2015Testing} as $T_{\rm CX}$ for simplifing notations. To distinguish the adaptive test with different $s_0$, we denote the adaptive test with any fixed $s_0$ as $TD^{N}_{s_0,{\rm ad}}$.

\begin{figure*}[hbpt]
	\includegraphics[width=16cm, height=18cm]{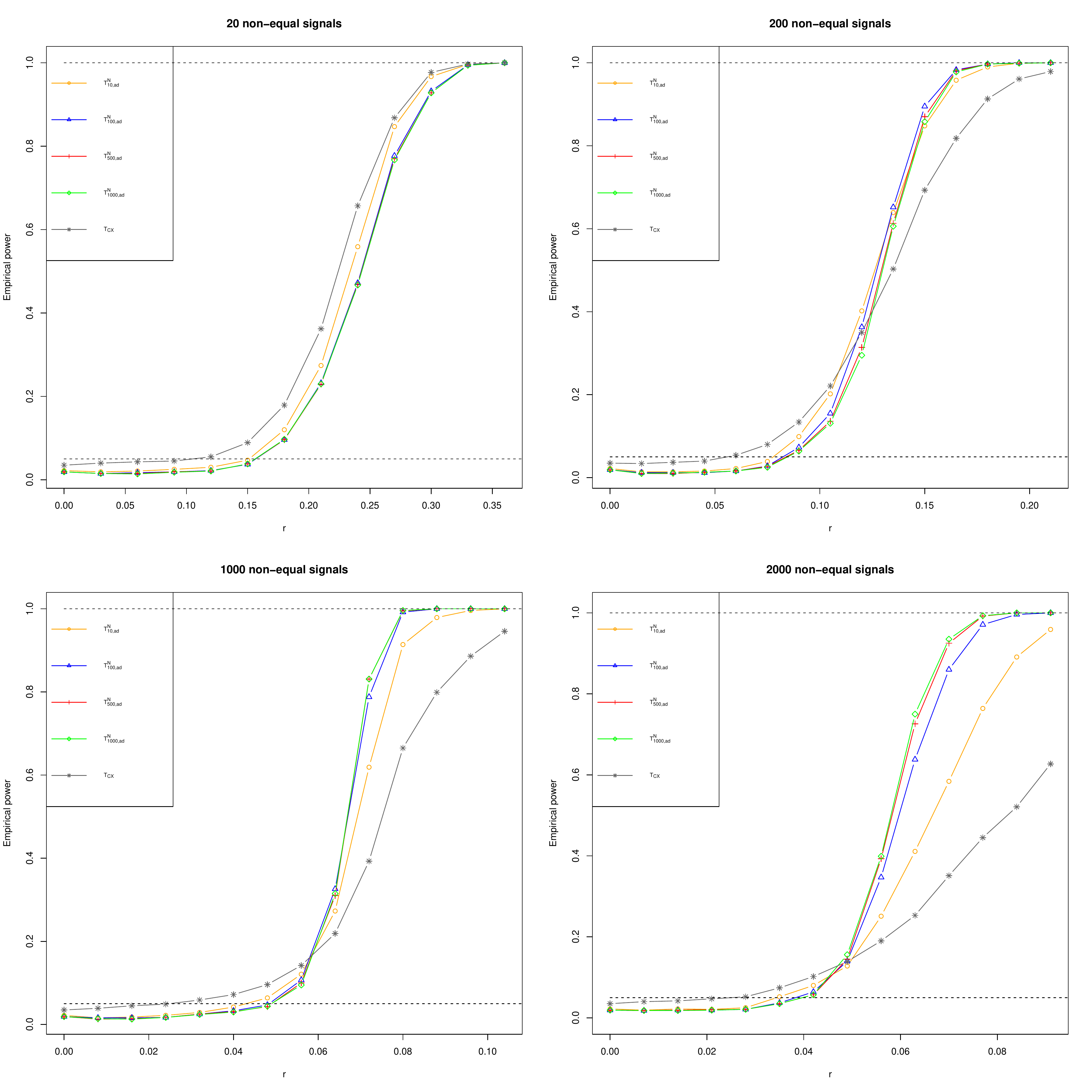}
	\caption{Empirical powers of various tests for {\bf Model 1}. The orange line with circles represents the adaptive test $T^{N}_{10,{\rm ad}}$, the blue line with triangles represents the adaptive test $T^{N}_{100,{\rm ad}}$, the red line with crosses represents the adaptive test $T^{N}_{500,{\rm ad}}$,  the green line with diamonds represents the adaptive test $T^{N}_{1000,{\rm ad}}$, the black line with stars represents the $T_{\rm CX}$ test.$\qquad \qquad \qquad \qquad \qquad \qquad \qquad \qquad \qquad \qquad \qquad \qquad \qquad \qquad \qquad \qquad \qquad$}\label{fig:model1}
\end{figure*}

In the simulation study, the sample sizes are set to be $n_1=n_2=200$, while the dimension $d=100$. Although the dimension $d$ seems to be small compared with the sample size, the parameters in the precision matrix which we are interested in are already much larger than the sample size ($d(d-1)/2$). In all simulations, the bootstrap sample sizes $B$ are set to be $1000$ and all the simulation results are based on $1000$ replications. Under the null hypothesis $\rm{\mathbf{H}_{0}}$, we set $\bOmega_2=\bOmega_1=\bOmega$. Under the alternative hypothesis $\rm{\mathbf{H}_{1}}$, we set $\bOmega_1=\bOmega+\delta \bI$ and  $\bOmega_2=\bOmega+\bGamma+\delta \bI$, where $\bGamma=(\gamma_{i,j})_{d\times d}$ is a nonzero matrix and $\delta=|\lambda_{\min}(\bOmega+\bGamma)|+0.05$. Suppose there are $m_t$ nonzero entries of $\bGamma$. Specially, we random sample $m_t/2$ locations in the upper triangle of $\bGamma$ and set each with a magnitude $r$. By the symmetric requirement of $\bOmega_2$, the location and magnitude of the other $m_{t}/2$ nonzero entries of $\bGamma$ can be determined by its upper triangle. To show that our test is adaptive to various alternative patterns, we set the nonzero entries of the $\bGamma$ as $m_t=20,200,1000,2000$. The $m_t=20,200$ are to illustrate the sparse alternatives and $m_t=1000,2000$ are to represent the dense alternatives. For all the simulations, simulation data are generated from multivariate Gaussian distributions with mean $\zero$ and covariance matrices $\bSigma_1=(\bOmega_1)^{-1}$ and $\bSigma_2=(\bOmega_2)^{-1}$. The nominal significance level for all the tests are set to be $\alpha=0.05$. To study the empirical performance of the test, the following three models of $\bOmega$ are considered.

\vspace{1em}
\textbf{Model 1:} $\bOmega^*=(\omega_{i,j}^{*})$ where $\omega_{i,i}^{*}=1$, $\omega_{i,j}^{*}=0.5\times \text{Bernoulli}(1,0.5)$ for $i<j$ and $\omega_{j,i}^{*}=\omega_{i,j}^{*}$. $\bOmega=(\bOmega^*+\delta \bI)/(1+\delta)$ with $\delta=|\lambda_{\min}(\bOmega^*)|+0.05$.

\vspace{1em}
\textbf{Model 2:} $\bSigma^*=(\sigma^{*(1)}_{i,j})$ where $\sigma^{*(1)}_{i,i}=1$, $\sigma^{*(1)}_{i,j}=0.5$ for $2(k-1)+1\leq i\neq j \leq 2k$, where $k=1,\ldots,[p/2]$ and $\sigma^{*(1)}_{i,j}=0$ otherwise. $\bOmega=\{(\bSigma^*+\delta \bI)/(1+\delta)\}^{-1}$ with $\delta=|\lambda_{\min}(\bSigma_1^*)|+0.05$.

\vspace{1em}
\textbf{Model 3:} $\bOmega^*=(\omega_{i,j}^{*(1)})$ where $\omega_{i,i}^{*(1)}=1$, $\omega_{i,j}^{*(1)}=0.5\times \text{Bernoulli}(1,0.3)$ for $i<j$ and $\omega_{j,i}^{*(1)}=\omega_{i,j}^{1}$. Other than that, we set $\omega_{i,j}^{*(1)}=\omega_{j,i}^{*(1)}=0.5$ for $i=20(k-1)+1$ and $20(k-1)+2\leq j \leq20(k-1)+20$, $1\leq k\leq p/20$. $\omega_{i,j}^{*(1)}=0$ otherwise. $\bOmega=(\bOmega^*+\delta \bI)/(1+\delta)$ with $\delta=|\lambda_{\min}(\bOmega^*)|+0.05$.
\vspace{1em}

The performances of the test methods under various alternative patterns for Model 1 are  shown  in Figure \ref{fig:model1}.
In Figure \ref{fig:model1}, the orange line with circles represents the adaptive test $T^{N}_{10,{\rm ad}}$, the blue line with triangles represents the adaptive test $T^{N}_{100,{\rm ad}}$, the red line with crosses represents the adaptive test $T^{N}_{500,{\rm ad}}$,  the green line with diamonds represents the adaptive test $T^{N}_{1000,{\rm ad}}$, the black line with stars represents the $T_{\rm CX}$ test proposed by \cite{Xia2015Testing}. The horizontal axis represents magnitude $r$ in the upper triangle of $\bGamma$,  a larger value of $r$ indicates a stronger signal. The vertical axis represents the empirical powers of different tests, while $r=0$ corresponds to the empirical sizes.

From Figure  \ref{fig:model1}, we can see that all the empirical sizes of different tests are under control. Under sparse alternative setting with $m_t=20$ (corresponds to the upper left panel in Figure \ref{fig:model1} with 20 non-equal signals), the empirical powers of maximum-norm based test $T_{\rm CX}$ are the highest and the empirical powers of the adaptive test with $s_0=10$ are still comparable though a little bit lower than CX test. Besides, with $s_0$ decreasing, the adaptive test tends to more powerful under sparse alternative setting. As the non-equal number $m_t$ becomes bigger, the empirical powers of the adaptive test with larger $s_0$  are getting better and better. Under the dense alternative (corresponds to the upper right panel and lower panels in Figure \ref{fig:model1} with more than 200 non-equal signals), the empirical powers of the adaptive tests are greater than those of the maximum-norm based CX tests with the magnitude $r$ larger than certain threshold. Although the empirical powers of adaptive test with different $s_0$ have some difference, the empirical powers of the adaptive test  show some robustness for the small changes of $s_0$. From Figure \ref{fig:model1}, we can also see that the empirical power of $T^{N}_{500,{\rm ad}}$ and $T^{N}_{1000,{\rm ad}}$ are almost equal to each other.

At last, we point out that the influence of the parameter $s_0$ on the power performance is more complicated. However,	by choosing $s_0$ close to half of the true number of nonzero signals $m_t/2$, the tests enjoy good performance. In practice, we can determine $s_0$ by the prior information or some empirical information.

The empirical results for Model $2$ and Model $3$ are similar as for Model 1 and thus are presented in the supplementary materials for saving space here.


\begin{figure}[!ht]
	\centering
	\includegraphics[width=12cm, height=8cm]{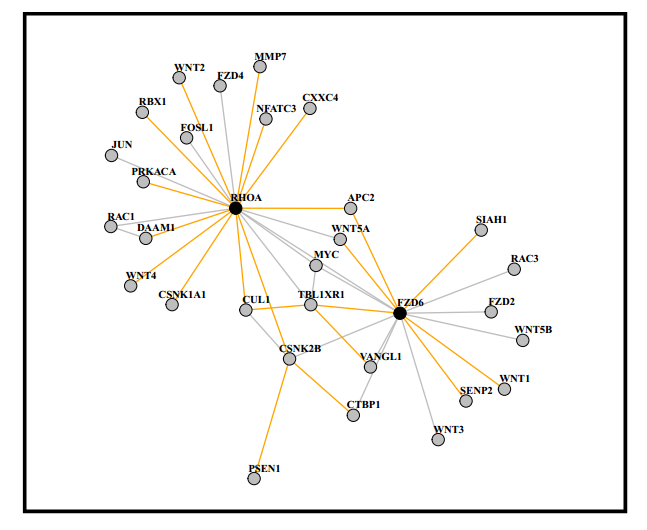}
	\caption{The differential networks estimated for the Wnt signaling pathway. Orange edges show an increase in conditional dependency from control group to lung cancer patient group; grey edges show a decrease.}\label{fig1}
\end{figure}

\subsection{Real data analysis}\label{subsec:realdata}

In this section, we apply our adaptive test method to a gene expression data set which is associated with lung cancer. The data set is publicly available from the Gene Expression Omnibus (\url{https://www.ncbi.nlm.nih.gov/geo/}) at accession number GDS2771. The data set is made up of 22,283 microarray-derived gene expression measurements from large airway epithelial cells sampled from 97 patients with lung cancer, and 90 control patients. \cite{mazieres2005wnt} showed that the Wnt pathway associated with lung cancer and many other lung diseases such as interstitial lung disease (ILD) and asthma. The Wnt pathway is also implicated in the development of several types of cancers, such as gastric cancer \citep{Clements2002Beta}, breast cancer \citep{Howe2004Wnt}.

Hence, in this paper, we focus our analysis on the $188$ genes in the Wnt signaling pathway, with $97$ patients with lung cancer and $90$ control patients. Gene expression levels were analyzed on a logarithmic scale and each gene feature was standardized within each group. Although the true conditional dependence relationships are unknown, we believe that there exists some specific links among genes in the Wnt signaling pathway of the patients with lung cancer. Hence, we use our method to test whether the underlying precision matrices of the patients with cancer or not are equal to each other. {In the real example, we know nothing about the underlying alternative patterns. Hence, as to the choice of $s_0$, we propose to tune $s_0$ in a finite set $\mathcal{S}$. Specifically, we set the doubly tuned data-adaptive test statistic as}
\begin{equation}\label{equ:adaptiveall}
TN_{\rm ad}= \min_{p \in \mathcal{P}, s_0 \in \mathcal{S}} \widehat{P}^{N}_{(s_0,p)}.
\end{equation}
{As long as the cardinality of the set $\mathcal{S}$ is fixed, all the theoretical properties for the adaptive test $N_{\rm ad}$ with fixed $s_0$  still hold for $TN_{\rm ad}$. Furthermore, the simulation study in Section \ref{subsec:simulation} showed that the empirical powers of the adaptive tests show robustness to different $s_0$. Hence, assuming  the cardinality of the set $\mathcal{S}$ to be finite is reasonable. Specifically, we choose $\mathcal{S}=\{10,50,100,500,1000,2000\}$ here. By the adaptive test $TN_{\rm ad}$, }
we reject the null hypothesis and think that there are difference for the underlying conditional dependence relationships. Hence, we use the differential network estimation approach in \cite{Zhao2015Direct} to estimate the differential network between the control group and the patient group. In detail, we choose the tuning parameter by the Bayesian information criterion (BIC) using the element-wise $L_{1}$ loss function. The differential network structure is given in Figure \ref{fig1}, in which the black edges represents the conditional correlations in lung cancer group are stronger compared with those in the control group, and the gray edges the other way around.  From Figure \ref{fig1}, many potentially important genes for lung cancer are detected, such as WNT1, WNT2, WNT5A etc, see \cite{mazieres2005wnt}. By Figure \ref{fig1}, we see that RHOA and FZD6 are two hubs in this graph. Hence, we may conclude that the connections of these two genes to other genes are important for identifying the lung cancer. Actually the importance of FDZ and RHOA can be referred to \cite{Corda2017Non}, \cite{Rapp2017WNT}.

\section{Discussion}\label{sec:discussion}
In this paper we propose an adaptive approach for testing the equality of two precision matrices, i.e. to investigate whether the network of connected node pairs change from one state to another. In the Gaussian  setting, the precision matrix can be described in terms of regression models and the elements of the precision matrix have a direct correspondence connection with the correlations of the error term. By Lasso or Dantzig selector, the regression coefficient estimator and the corresponding estimated regression errors are obtained. Based on the bias corrected estimator of the correlations of the error terms, we propose to construct a family of $(s_0,p)$-norm based test statistics with different $p$. By taking the minimum $P$-value of these tests, we construct an adaptive test statistics.  We utilize multiplier bootstrap method to approximate the limiting distribution of the test statistic. Theoretical guarantees are provided for the proposed procedure and numerical study illustrates its good empirical performance under various alternatives.

The current work relies heavily on the Gaussian graph assumption which is sometimes restrictive in real application. In the future, we will consider the adaptive test of more general graphical models.

\section*{Acknowledgements}
Yong He's research is partially supported by the grant of the National Science Foundation of China (NSFC 11801316) and National Statistical Scientific Research Project (2018LY63).  Xinsheng Zhang's research is partially supported by the grant of the National Science Foundation of China (NSFC 11571080).
		
\bibliographystyle{plain}
\bibliography{KaiTiRef}

\clearpage

\section*{APPENDIX SECTION}
\begin{appendix}
\section{Technical lemmas}
Before proving the main results, we introduce some technical lemmas which is useful in proving the main theorems. Let $\bZ_1, \ldots, \bZ_n$ be independent centered random vectors in $\mathbb{R}^{d}$ with $\bZ_{k}=(Z_{k1},\ldots, Z_{kd})$, for $k=1,\ldots,n$ and $\bG_{k}$ ($k=1,\ldots,n$) be independent Gaussian random vectors in $\mathbb{R}^{d}$ with the same mean vector and covariance matrix as $\bZ_{k}$, and assume the following conditions hold:

\vspace{0.8em}

{\bf  (M1)} $n^{-1}\sum_{k=1}^n\E\big[(\vb'\bZ_{k})^2\big]\ge b>0$ for any $\vb\in \mathcal{V}_{s_0}$ with $\mathcal{V}_{s_0}:=\{\vb\in \mathbb{S}^{d-1}: \|\vb\|_0\le s_0\}$.
\vspace{0.8em}

{\bf (M2)} $n^{-1}\sum_{k=1}^n\E\big[|Z_{ki}|^{2+\ell}\big]\le Q^\ell$  for $\ell=1,2$ and $i=1,\ldots,d$.
\vspace{0.8em}

{\bf (M3)} $\E\big[\exp(|Z_{ki}|/Q)\big]\le 2$ for $i=1,\ldots,d$ and $k=1,\ldots,n$.
\vspace{0.8em}

\begin{lemma}\label{lemma:CLT} (Lemma A.1 in \cite{Zhou2018Unified})
	Assume $s_0^2\log(dn)=O(n^\zeta)$ with $0<\zeta<1/7$ and $\bZ_1, \ldots,\bZ_{n}$ satisfy {\bf (M1)}, {\bf (M2)}, and {\bf (M3)}. For $1\le p\le \infty$ and sufficiently large $n$, there is a constant $\zeta_0>0$ such that
	\begin{equation*}
	\begin{array}{ll}
	\sup_{z\in (0,\infty)}\Big|\P\Big(&\|S_{n}^{\bZ}\|_{(s_0,p)}\le z\Big)-\P\Big(\|S_{n}^{\bG}\|_{(s_0,p)}\le z\Big)\Big|
	\le {n^{-\zeta_0}},
	\end{array}
	\end{equation*}
	where $S_{n}^{\bZ}=n^{-1/2}\sum\nolimits_{k=1}^{n}\bZ_{k}$, $S_{n}^{\bG}=n^{-1/2}\sum\nolimits_{k=1}^{n}\bG_{k}$ and  $C$  depends on $b$ and $Q$.
\end{lemma}

\begin{lemma}(Product of sub-Gaussian is sub-exponential, Lemma 2.7.7 in \cite{Roman2017High})\label{lemma:subgaussian}
	{}			Let $X$ and $Y$ be sub-Gaussian random variables, we have $XY$ is sub-exponential. Moreover,
	\begin{equation*}
	\|XY\|_{\psi_{1}}\le \|X\|_{\psi_{2}}\|Y\|_{\psi_{2}},
	\end{equation*}
	where the $\psi_{\alpha}$-norm ($\alpha \ge 1$) of $X$ is defined by
	\begin{equation*}
	\|X\|_{\psi_{\alpha}} := {\rm inf}\Big(c>0: \E\big(\exp(|X|^{\alpha}/c^{\alpha})\big)\le 2\Big).
	\end{equation*}
\end{lemma}

We then introduce the following lemma to get the  bound for $U_{i,j,m}$ and the bias of $\hat{r}_{i,j,m}$ uniformly in $1\le i<j\le d$.	
\begin{lemma}\label{lemma:Uij}
	With Assumption \textbf{(B)-(D)} hold, and for $\log d=\bo\{n^{1/2}\}$, we have
	\begin{equation}\label{equ:Uij}
	\max_{1\le i<j\le d} |U_{i,j,m}|=\bOp\{(\log d/n)^{1/2}\},
	\end{equation}
	and
	\begin{equation}\label{equ:hatrij}
	\max_{1\le i<j\le d}|r_{i,j,m}-\hat{r}_{i,j,m}|=\bOp\{(\log d/n)^{1/2}\},
	\end{equation}
	for sufficient large $n$.
\end{lemma}
The proof of this lemma is in Section 1.2 of the supplementary materials.

Corresponding to the definition of $\bSigma_{m}^{\tilde{U}}$ in Section \ref{subsec::assumption}, we introduce its plug-in covariance matrix estimator as $\hat{\bSigma}_{m}^{\tilde{U}}=(\hat{\sigma}^{\tilde{U}}_{s,t,m})_{1\le s,t\le d(d-1)/2}$, where
\begin{equation}\label{est:covU}
\hat{\sigma}^{\tilde{U}}_{s,t,m} = \left\{\
\begin{aligned}
\hat{\theta}_{i,j,m}=
\frac{1+\hat{\beta}_{i,j,m}^{2}\hat{r}_{i,i,m}/\hat{r}_{j,j,m}}{n_m\hat{r}_{i,i,m}\hat{r}_{j,j,m}},  s=t,\\
\frac{\hat{r}_{i_{1},i_{2},m}\hat{r}_{j_{1},j_{2},m}+\hat{r}_{i_{1 }, j_{2},m}\hat{r}_{i_{2},j_{1},m}}{n_m\hat{r}_{i_1,i_1,m}\hat{r}_{j_1,j_1,m}\hat{r}_{i_2,i_2,m}\hat{r}_{j_2,j_2,m}}, s\neq t,
\end{aligned}
\right.
\end{equation}
with $1\le i<j\le d$, $1\le i_1<j_1\le d$, $1\le i_2<j_2\le d$, $i_1\neq i_2$, $j_1\neq j_2$, $m=1,2$.

Set the correlation matrix of $\text{trivec}(\tilde{\Ub}_{m})$ as $\mathbf{R}^{\tilde{U}}_{m}=(r^{\tilde{U}}_{s,t,m})_{1\le s, t\le d(d-1)/2}$, and its plug-in estimator as $\hat{\mathbf{R}}^{\tilde{U}}_{m}=(\hat{r}^{\tilde{U}}_{s,t,m})_{1\le s,t\le d(d-1)/2}$. By the definition of correlation, we have
\begin{equation*}
r^{\tilde{U}}_{s,t,m}=\frac{\sigma^{\tilde{U}}_{s,t,m}}{\sigma^{\tilde{U}}_{s,s,m}\sigma^{\tilde{U}}_{t,t,m}}, \ \
\hat{r}^{\tilde{U}}_{s,t,m}=\frac{\hat{\sigma}^{\tilde{U}}_{s,t,m}}{\hat{\sigma}^{\tilde{U}}_{s,s,m} \hat{\sigma}^{\tilde{U}}_{t,t,m}}.
\end{equation*}
With the Lemma \ref{lemma:Uij} holding, we then introduce the following lemma to bound the estimation error of the plug-in estimator $\hat{\sigma}^{\tilde{U}}_{s,t,m}$ and $\hat{r}_{s,t}$.

\begin{lemma}\label{lemma:bigD}
	With assumptions  {\bf (B)}-{\bf(D)} hold, and for $\log d=\bo\{n^{1/2}\}$, we have
	\begin{equation}\label{equ:plugin}
	\begin{aligned}
	&\max_{1\le s,t \le d(d-1)/2\atop m=1,2}\Big(\big|\hat{\sigma}^{\tilde{U}}_{s,t,m}-\sigma^{\tilde{U}}_{s,t,m}\big|,\big|\hat{r}^{\tilde{U}}_{s,t,m}-r^{\tilde{U}}_{s,t,m}\big|\Big)=\bOp\{(\log d/n)^{1/2}\},
	\end{aligned}
	\end{equation}
	for sufficient large $n$.
\end{lemma}
The proof of Lemma \ref{lemma:bigD} is in Section 1.3 of the supplementary materials.

Lemma \ref{lemma:bigD} bound the estimation error of the plug-in estimator, it is important to get the approximated distribution of $\|{\rm trivec}(\Wb)\|_{(s_0,p)}$, i.e. the proof of Theorem \ref{theorem:bootstrap}. In addition, to prove Theorem \ref{theorem:bootstrap}, the approximated distribution for $\|{\rm trivec}(\Wb^{b})\|_{(s_0lp)}$ is also needed. By the definition of $T^{b}_{i,j,m}$ and $W^{b}_{i,j}$in \eqref{Wbootsineqj}, and as $\eta^{b}_{m,k}, m=1,2$, $k=1,\ldots,n$ are independent standard normal random variables,
we have $T^{b}_{i,j,m}|\mathcal{X},\mathcal{Y}\sim \bN(0,\tilde{\theta}_{i,j,m})$ with $1\le i,j\le d$.
Set $\Tb^{b}_{m}|\mathcal{X},\mathcal{Y}=(T^{b}_{i,j,m}|\mathcal{X},\mathcal{Y})^{\top}_{d\times d}$, we then have $\text{trivec}(\Tb^{b}_{m}|\mathcal{X},\mathcal{Y})
\sim \bN(0,\hat{\bSigma}^{\bT^{b}}_{m})$,
where $\hat{\bSigma}^{\Tb^{b}}_{m}=\big( \hat{\sigma}^{\Tb^{b}}_{s,t,m}\big)_{1\le s,t\le d(d-1)/2}$ and
$\hat{\sigma}^{\Tb^{b}}_{s,t,m}=\tilde{\theta}_{i,j,m}$ for $s=t$. Actually, it also can be seen as the sample estimator for $\bSigma_{m}^{\tilde{U}}$.
Similarly,  set the corresponding correlation estimator as $\widehat{\mathbf{R}}^{\Tb^{b}}_{m}=(\hat{r}^{\Tb^{b}}_{s,t,m})_{1\le s, t\le d(d-1)/2}$,
with $\hat{r}^{\tilde{U}}_{s,t,m}={\hat{\sigma}^{\Tb^{b}}_{s,t,m}}/{(\hat{\sigma}^{\Tb^{b}}_{s,s,m} \hat{\sigma}^{\Tb^{b}}_{t,t,m})}$. We then provide the following lemma to bound the estimation errors.
\begin{lemma}\label{lemma:bootsigma}
	With Assumption \text{(B)-(D)} hold, and for $\log d=\bo\{n^{1/2}\}$, we have
	\begin{equation}\label{equ:bootsample}
	\begin{aligned}
	&\max_{1\le s,t \le d(d-1)/2\atop m=1,2}\Big(\big|\hat{\sigma}^{\Tb^{b}}_{s,t,m}-\sigma^{\tilde{U}}_{s,t,m}\big|, \big|\hat{r}^{\Tb^{b}}_{s,t,m}-r^{\tilde{U}}_{s,t,m}\big|\Big)=\bOp\{(\log d/n_{m})^{1/2}\}.
	\end{aligned}
	\end{equation}
\end{lemma}
The proof of Lemma \ref{lemma:bootsigma} is in the supplementary materials.

\section{Proof of Theorems}

{By the results in Lemma \ref{lemma:bigD}, Lemma \ref{lemma:bootsigma}, and Theorem 4.1,  actually the proof of Theorem 4.2 is similar to the proof of Theorem 3.6  in \cite{Zhou2018Unified},  the proof of Theorem 4.3 is similar to the proof of Theorem 3.3 and Theorem 3.7  in \cite{Zhou2018Unified}. Hence, we omit these proofs and only show the detailed proof of Theorem \ref{theorem:bootstrap}.}

\subsection{Proof of Theorem \ref{theorem:bootstrap}}
\begin{proof}
	We provide the detail proof of (\ref{equ:bootstrap}) in two steps.
	
	\textbf{Step (i).}
	In this step, we establish the approximated distribution of ${\rm trivec}(\Wb)$. To this end, we introduce another intermediate variable. Define $\Hb \in \reals^{d\times d}$ with
	\begin{equation*}
	H_{i,j}=(\tilde{U}_{i,j,1}-\tilde{U}_{i,j,2})/\sqrt{\theta_{i,j,1}+\theta_{i,j,2}}, \quad 1 \le i,j \le d.
	\end{equation*}
	
	The following lemma establish that ${\rm trivec}(\Hb)$  is a good approximation of ${\rm trivec}(\Wb)$ under $(s_0,p)$-norm.
	\begin{lemma}\label{lemmaWH}
		We assume that Assumption {\textbf{(A)}-\textbf{(D)}} hold. Under $\Hb_0$ of (\ref{hyp:omega2test0}), we have that there is a constant $C>0$ such that
		\begin{equation*}
		\P\left(\|{\rm trivec}(\Wb)-{\rm trivec}(\Hb)\|_{(s_0,p)}>\epsilon\right)= \bo(1),
		\end{equation*}
		as $n\rightarrow\infty$, where $\epsilon= \bO\big\{s_0(\log d/n)^{1/2}\big\}$.
	\end{lemma}
	The proof for this lemma is in the supplementary materials.
	By the definition of $\tilde{U}_{i,j,m}$ and by Lemma \ref{lemma:subgaussian}, the Assumptions \textbf{(M1)-(M3)} are hold.
	As ${\rm trivec}(\Hb)$ is a sum of independent random vectors with mean zero and covariance matrix $\mathbf{R}^{\tilde{U}}_{12}$, where $\mathbf{R}^{\tilde{U}}_{12}=(D^{\tilde{U}}_{12})^{-1/2}\bSigma^{\tilde{U}}_{12}(D^{\tilde{U}}_{12})^{-1/2}$ with $\bSigma^{\tilde{U}}_{12}=\bSigma^{\tilde{U}}_{1}+\bSigma^{\tilde{U}}_{2}$ and $D^{\tilde{U}}_{12}=\text{Diag}(\bSigma^{\tilde{U}}_{12})$. we use a Gaussian random vector with the same covariance matrices as its approximation. Let $\bm{G}\in \reals^{d(d-1)/2}$ be a Gaussian random vector with mean zero and covariance matrix $\mathbf{R}^{\tilde{U}}_{12}$.
	By Lemma \ref{lemma:CLT}, we have
	\begin{equation*}
	\begin{split}
	\mathop{\mathrm{sup}}\limits_{z-\epsilon >0}\Big|\P(\|{\rm trivec}(\Hb)\|_{(s_0,p)}>z-\epsilon)-\P(\|\bm{G}\|_{(s_0,p)}>z-\epsilon)\Big| \leq Cn^{-\zeta_{0}}.
	\end{split}
	\end{equation*}
	and
	\begin{equation*}
	\begin{split}
	\mathop{\mathrm{sup}}\limits_{z+\epsilon >0}\Big|\P(\|{\rm trivec}(\Hb)\|_{(s_0,p)}>z+\epsilon)-\P(\|\bm{G}\|_{(s_0,p)}>z+\epsilon)\Big| \leq Cn^{-\zeta_{0}}.
	\end{split}
	\end{equation*}
	Further, by the triangle inequality, we have
	\begin{equation}\label{absT*1}
	\begin{split}
	\P(\|{\rm trivec}(\Wb)\|_{(s_0,p)}>z)\le \P(\|{\rm trivec}(\Hb)\|_{(s_0,p)}>z-\epsilon)
	+\P(\|{\rm trivec}(\Wb)-{\rm trivec}(\Hb)\|_{(s_0,p)}>\epsilon)
	\end{split}
	\end{equation}
	\begin{equation}\label{absT*2}
	\begin{split}
	\P(\|{\rm trivec}(\Wb)\|_{(s_0,p)}>z)\ge \P(\|{\rm trivec}(\Hb)\|_{(s_0,p)}>z+\epsilon)
	-\P(\|{\rm trivec}(\Wb)-{\rm trivec}(\Hb)\|_{(s_0,p)}>\epsilon).
	\end{split}
	\end{equation}

	Thus, by the triangle inequality and  Lemma \ref{lemmaWH}, and combining Equation \eqref{absT*1} and \eqref{absT*2}, we have
	\begin{equation}\label{equ:trivWG}
	\begin{aligned}
	\P\Big(\|\bG\|_{(s_0,p)}>z+\epsilon\Big)-\bo(1)
	\le\P\Big(\|{\rm trivec}(\Wb)\|_{(s_0,p)}>z\Big)
	\le\P\Big(\|\bG\|_{(s_0,p)}>z-\epsilon\Big)+\bo(1).
	\end{aligned}
	\end{equation}
	
	\begin{lemma}\label{lemma:Gz-ez}(Lemma B.2 in \cite{Zhou2018Unified})
		Assumptions {\bf (A)} hold. For any $z>0$ and $\varepsilon=\bO\big(s_{0}\log^{2}(dn)n^{-1/2}\big)$, we have $\P\Big(z-\varepsilon<\|\bG\|_{(s_0,p)}\le z\Big)=\bo(1)$ as $n\rightarrow\infty$.
	\end{lemma}		
	By Lemma \ref{lemma:Gz-ez} and combining \eqref{equ:trivWG}, as $n\rightarrow \infty$, we have
	\begin{equation}\label{T*G}
	\begin{split}
	\mathop{\mathrm{sup}}\limits_{z>0}\Big|\P\big(&\|{\rm trivec}(\Wb)\|_{(s_0,p)}>z\big)-\P\big(\|\bG\|_{(s_0,p)}>z\big)\Big|=\bo(1).
	\end{split}
	\end{equation}
	
	\textbf{Step (ii).} In this step, we aim to obtain the distribution of ${\rm trivec}(\Wb^{b})$ given $\mathcal{X}$ and $\mathcal{Y}$. By the definition of $T^{b}_{i,j,m}$in \eqref{Wbootsineqj}, as $\eta^{b}_{m,k}, m=1,2$, $k=1,\ldots,n$ are independent standard normal random variables, we have $T^{b}_{i,j,m}|\mathcal{X},\mathcal{Y}\sim \bN(0,\tilde{\theta}_{i,j,m})$ with $1\le i,j\le d$, and $\text{trivec}(\Tb^{b}_{m}|\mathcal{X},\mathcal{Y})
	\sim \bN(0,\hat{\bSigma}^{\bT^{b}}_{m})$ with $\Tb^{b}_{m}|\mathcal{X},\mathcal{Y}=(T^{b}_{i,j,m}|\mathcal{X},\mathcal{Y})^{\top}_{d\times d}$. By setting $\widehat{\bSigma}^{\bT^{b}}_{12}=\widehat{\bSigma}^{\bT^{b}}_1/n_1+\widehat{\bSigma}^{\bT^{b}}_2/n_2$ and $\widehat{\mathbf{D}}^{\bT^{b}}_{12}=\mathrm{Diag}(\widehat{\bSigma}^{\bT^{b}}_{12})$, and conditional on $\mathcal{X}$ and $\mathcal{Y}$, we have
	\[
	{\rm trivec}({\Wb^{b}})\sim N(\zero,\widehat{\mathbf{R}}^{\bT^{b}}_{12}),
	\]
	where $\widehat{\mathbf{R}}^{\bT^{b}}_{12}:=(\widehat{\mathbf{D}}_{12}^{\bT^{b}})^{-1/2}\widehat{\bSigma}^{\bT^{b}}_{12}(\widehat{\mathbf{D}}_{12}^{\bT^{b}})^{-1/2}$.
	
	Recall that in last step, we have  $\bG\sim N(\mathbf{0},\bR^{\tilde{U}}_{12})$. 
	By Lemma \ref{lemma:bootsigma} and  similar argument in Lemma B.3 of \cite{Zhou2018Unified}, we get the following lemma to establish the upper bound  for the approximation error between $\|\bm{T}^{*b}\|_{(s_0,p)}$ and $\|\bm{G}\|_{(s_0,p)}$.
	\begin{lemma}\label{lemma:GT*b}
		Under Assumptions \textbf{(A)-(D)}, with probability at least $1-Cn^{-1}$, we have
		\[
		\begin{array}{ll}
		{\mathop{\mathrm{sup}} \limits_{z>0}\Big|\mathbb{P}\big(\|\bm{G}\|_{(s_0,p)}> z\big)}
		-\mathbb{P}\big(\|{\rm trivec}(\bm{W}^{b})\|_{(s_0,p)}> z|\mathcal{X},\mathcal{Y}\big)\Big|
		=\bo(1).
		\end{array}
		\]
	\end{lemma}
	
	By the triangle inequality, we have
	\begin{equation*}
	\begin{array}{lll}
	\left|\mathbb{P}(N_{(s_0,p)}> z)-\mathbb{P}(N^{b}_{(s_0,p)}>z|\mathcal{X},\mathcal{Y})\right|\le&
	\Big|\mathbb{P}(\|{\rm trivec}(\bm{W})\|_{(s_0,p)}> z)-\mathbb{P}(\|\bG\|_{(s_0,p)}>z)\Big|\\
	&+\Big|\mathbb{P}(\|\bG\|_{(s_0,p)}\!>z)\!-\!\mathbb{P}(\|{\rm trivec}(\bm{W}^{b})\|_{(s_0,p)}\!>z|\mathcal{X},\mathcal{Y})\Big|.
	\end{array}
	\end{equation*}
	Hence, combining Equation \eqref{T*G} and the result in Lemma \ref{lemma:GT*b},  we have
	\[
	\mathop{\mathrm{sup}}\limits_{z \!\in\!(0,\infty)}\left|\mathbb{P}\big(N_{(s_0,p)}\!>\!z\big)\!-\!\mathbb{P}\big(N^{b}_{(s_0,p)}\!>\!z|\mathcal{X},\mathcal{Y}\big)\!\right|\!\!=\!\bo(1),
	\]
	which finishes the proof of Theorem \ref{theorem:bootstrap}.
	
\end{proof}

	\section{Proof of Lemmas}\label{sec:prooflemmas}
	Before presenting the detailed proofs for the useful lemmas in the Appendix of the main paper, we introduce some additional useful lemmas.
	
	\subsection{Some useful lemmas}\label{sec::prooflemma}
	
	\begin{lemma}\label{lemma:inverse} ( Lemma C.1 in \cite{Zhou2018Unified})
		$\xi_1, \ldots,\xi_q\in \mathbb{R}$ are positive random variables. For $y\in (0,1]$, we have
		\begin{equation}\label{ineq:inverseRelationship}
		\P\Big(\max_{1\le s\le q}|1-\xi_s|\le y/2\Big)\le
		\P\Big(\max_{1\le s\le q}|1-\xi_s^{-1}|\le y\Big).
		\end{equation}
	\end{lemma}
	
	Next, we introduce another very useful lemma proposed by \cite{Smith1918ON}, which allows one to compute higher-order moments of the multivariate normal distribution in terms of its covariance matrix.
	
	\begin{lemma}\label{lemma:Bernsteinsubsub}(Bernstein Inequality for sub-exponential random vector) Let $X_{1}, \ldots, X_{n}$ be independent, mean zero, sub-exponential random variables. There exists an absolute constant $c>0$  such that
		\begin{equation*}
		\P\Big(\frac{1}{n}\sum_{i=1}^{n}X_{i} \ge t \Big)< 2\exp\left(-cn\min\Big(\frac{t^2}{\bar{v}},\frac{t}{M}\Big)\right),
		\end{equation*}
		hold for any $t>0$, where $\bar{v}=n^{-1}\sum_{i=1}^{n}\|X_{i}\|^{2}_{\psi_{1}}$, $M=\max_{i}\|X_{i}\|_{\psi_{1}}$.
	\end{lemma}
	
	Next, we introduce another very useful lemma which allows one to compute higher-order moments of the multivariate normal distribution in terms of its covariance matrix.
	\begin{lemma}(Isserlis' Theorem)\label{lemma:Isserlistheorem}
		Let $(Z_1,Z_2,Z_3,Z_4)^{\top}$ be a zero-mean multivariate normal random vector, we  have
		\begin{equation*}
		\mathbf{E}(Z_1Z_2Z_3Z_4)=\mathbf{E}(Z_1Z_2)\mathbf{E}(Z_3Z_4)+\mathbf{E}(Z_1Z_3)\mathbf{E}(Z_2Z_4)+\mathbf{E}(Z_1Z_4)\mathbf{E}(Z_2Z_3).
		\end{equation*}
	\end{lemma}

	\subsection{Proof of Lemma A.3}\label{proof:LemmaA.3}
	\begin{proof}
		\textbf{(i)} In this part, we show the proof for (A.1).
		By the setting in (2.1), $\epsilon_{k,i,m}$ is following the normal distribution with mean zero and variance $\sigma_{i,i,m}-\bSigma_{i,-i,m}\{\bSigma_{-i,-i,m}\}^{-1}\bSigma_{-i,i,m}$.  As we know, the centered Gaussian random variable is also sub-Gaussian distributed. 	Hence, by Lemma A.2, we get $\epsilon_{k,i,m}\epsilon_{k,j,m}$ is following sub-exponential distribution. By the Bernstein inequality, we have
		\begin{equation}\label{equ:bernsteinU}
		\P\Bigg(\bigg|\frac{1}{n_{m}} \sum_{k=1}^{n_m} \Big\{ \epsilon_{k,i,m} \epsilon_{k,j,m} -\mathbb{E}\big(\epsilon_{k,i,m}\epsilon_{k,j,m}\big) \Big\}\bigg|\ge t\Bigg)\le 2\exp \Big(-cn\min\big(c_1{t^{2}}, c_2{t} \big) \Big),
		\end{equation}
		where $c,c_1,c_2$ are positive constants. By the inequality of
		\[
		\P\Big(\max_{1\le i<j\le d}|U_{i,j,m}|\ge t\Big)\le \frac{d(d-1)}{2}\P\Big(|U_{i,j,m}|\ge t\Big),
		\]
		and combining the Bernstein inequality in \eqref{equ:bernsteinU},
		we have $\max_{1\le i<j\le d}|U_{i,j,m}|=\bO_{p}\{(\log d/n)^{1/2}\}$, which finishes the proof of (A.1).
		
		\textbf{(ii)} In this part, we aim to prove (A.2).
		By the Lemma A.2 of \cite{Xia2015Testing}, we have
		\begin{equation}\label{equ:rU}
		\hat{r}_{i,j,m}-(w_{i,i,m}\hat{\sigma}_{i,i,m,\epsilon}+w_{j,j,m}\hat{\sigma}_{j,j,m,\epsilon}-1)r_{i,j,m}=-U_{i,j,m}+\bo_{p}\{(n_m\log d)^{-1/2}\},
		\end{equation}
		uniformly in $1\le i<j\le d$.
		Noted that $\max \limits_{1\le i<j\le d}|w_{i,i,m}\hat{\sigma}_{i,i,m,\epsilon}+w_{j,j,m}\hat{\sigma}_{j,j,m,\epsilon}-2|=\bO_{p}\{(\log d/n)^{1/2}\}$,  where $\hat{\sigma}_{i,j,m,\epsilon}=(1/n_m)\sum_{k=1}^{n_m}(\epsilon_{k,i,m}-\bar{\epsilon}_{i,m})(\epsilon_{k,j,m}-\bar{\epsilon}_{j,m})$ with $\bar{\epsilon}_{i,m}=(1/n_m)\sum_{k=1}^{n_m}\epsilon_{k,i,m}$. By the triangle inequality,  and combining \eqref{equ:rU} and (A.1), we then get
		\begin{equation*}
		\max_{1\le i<j\le d}|\hat{r}_{i,j,m}-r_{i,j,m}|\le \bO_{p}\{\max _{1\le i<j \le d}|r_{i,j,m}|(\log d/n)^{1/2}\}+\max_{1\le i<j\le m}|U_{i,j,m}|+\bo_{p}\{(n_m\log d)^{-1/2}\}.
		\end{equation*}
		Hence, by Assumption  \textbf{(B)}, we have $\max \limits_{1\le i<j\le d}|\hat{r}_{i,j,m}-r_{i,j,m}|= \bO_{p}\{(\log d/n)^{1/2}\}$, which finishes the proof of (A.2).
	\end{proof}
	
	\subsection{Proof of Lemma A.4}
	\begin{proof}
		Due to the expression of $\sigma_{s,t,m}$ are different for $s, t$, see (4.1), We prove Lemma A.4 in two steps. In the first step, we bounded the estimation of  $\hat{\sigma}_{s,t,m}$  for $s=t$. In the second step, we bound the estimation of  $\hat{\sigma}_{s,t,m}$ for $s\neq t$.
		
		\textbf{Step 1.} For $s=t$, by the definition of $\hat{\sigma}_{s,s,m}$ and $\sigma_{s,s,m}$ in (A.3) and (4.1), we have
		\begin{equation*}
		|\hat{\sigma}_{s,s,m}-\sigma_{s,s,m}|= \bigg|\frac{\hat{\beta}_{i,j,m}^{2}}{n_m\hat{r}^{2}_{j,j,m}}-
		\frac{\beta_{i,j,m}^{2}}{n_mr^{2}_{j,j,m}}+\frac{1}{n_m\hat{r}_{i,i,m}\hat{r}_{j,j,m}}-\frac{1}{n_mr_{i,i,m}r_{j,j,m}}\bigg|.	
		\end{equation*}
		Considering the triangle inequality and $r_{j,j,m}$ is positive, we get
		\begin{equation}\label{equ:thetaijhat}
		\begin{split}
		|\hat{\sigma}_{s,s,m}-\sigma_{s,s,m}|=|\hat{\theta}_{i,j,m}-&\theta_{i,j,m}|\le \underbrace{\frac{\big|\hat{\beta}_{i,j,m}-
				\beta_{i,j,m}\big|\big(2|\beta_{i,j,m}|+\big|\hat{\beta}_{i,j,m}-\beta_{i,j,m}\big|\big)}{n_mr^{2}_{j,j,m}}\bigg|\frac{r^{2}_{j,j,m}}{\hat{r}^{2}_{j,j,m}}\bigg|}_{L_{i,j,m}}\\
		&+\underbrace{\frac{\beta^{2}_{i,j,m}+(\beta_{i,j,m}-\hat{\beta}_{i,j,m})^2+2|\beta_{i,j,m}||\beta_{i,j,m}-\hat{\beta}_{i,j,m}|}{n_mr^{2}_{j,j,m}}\bigg|\frac{r^{2}_{j,j,m}}{\hat{r}^{2}_{j,j,m}}-1\bigg|}_{L^{'}_{i,j,m}}\\
		&+\underbrace{\frac{1}{n_mr_{i,i,m}r_{j,j,m}}\bigg|\frac{r_{i,i,m}r_{j,j,m}}{\hat{r}_{i,i,m}\hat{r}_{j,j,m}}-1\bigg|}_{L^{''}_{i,j,m}}.
		\end{split}
		\end{equation}
		By the Lemma A.2 in \cite{Xia2015Testing}, we have
		\begin{equation}\label{equ::maxriii}
		\max \limits_{1\le i \le d}\big|\hat{r}_{i,i,m}-r_{i,i,m}\big|=\bO_{p}\{(\log d/n_{m})^{1/2}\}.
		\end{equation}
		With the sparsity conditon in Assumption \textbf{(D)} and combining Assumption \textbf{(B)}, the Equation (2.2) hold. Furthermore, we obtain
		\begin{equation}\label{equ::maxbeta}
		\bmax \limits_{1\leq i\le d, 1\le j\le d-1}
		\big|\hat{\beta}_{i,j,m}-\beta_{i,j,m}\big|_1=\bo_{p}\big\{(\log d)^{-1}\big\}.
		\end{equation}
		Considering $\beta_{i,j,m}=-w_{i,j,m}/w_{j,j,m}$, $w_{i,j,m}=r_{i,j,m}/(r_{i,i,m}r_{j,j,m})$ and $w_{j,j,m}=1/r_{j,j,m}$, we obtain that $\beta_{i,j,m}=-\frac{r_{i,j,m}}{r_{i,i,m}}$. By Assumption \textbf{(B)}, $|\beta_{i,j,m}|$ is bounded.
		by Assumption \textbf{(B)} and the bound in \eqref{equ::maxriii}, there exist an constant $c_0$, such that
		\begin{equation}\label{equ:hatrrjj2}
		\max_{1\le j\le d}\big|\hat{r}^{2}_{j,j,m}/r^{2}_{j,j,m}-1\big|=2\max_{1\le j\le d}\big(w_{j,j,m}\big|\hat{r}_{j,j,m}-r_{j,j,m}\big|\big)=\bO_{p}\{(\log d/n_{m})^{1/2}\}.
		\end{equation}
		By Lemma \ref{lemma:inverse}, and combining \eqref{equ::maxriii} and \eqref{equ::maxbeta}, we have
		\begin{equation}\label{equ:M'}
		\max_{1\le i<j\le d}L_{i,j,m}=\bO_{p}\{(n_{m}\log d)^{-1}\}.
		\end{equation}
		For $L^{'}_{i,j,m}$, by Lemma \ref{lemma:inverse} and combining \eqref{equ:hatrrjj2}, (2.2) and Assumption \textbf{(B)},  we have
		\begin{equation}\label{equ:M}
		\max_{1\le i<j\le d}L^{'}_{i,j,m}=\bO_{p}\{(\log d/n_{m})^{1/2}\}.
		\end{equation}
		For $L^{''}_{i,j,m}$, by triangle inequality, Assumption \textbf{(B)} and \eqref{equ::maxriii}, we have
		\begin{equation*}
		\begin{split}
		\bigg|\frac{\hat{r}_{i,i,m}\hat{r}_{j,j,m}}{r_{i,i,m}r_{j,j,m}}-1\bigg|&\le \frac{\big|(r_{i,i,m}-\hat{r}_{i,i,m})\big|\big|(r_{j,j,m}-\hat{r}_{j,j,m})\big|}{r_{i,i,m}r_{j,j,m}}+\frac{\big|r_{j,j,m}-\hat{r}_{j,j,m}\big|}{r_{j,j,m}}+\frac{\big|r_{i,i,m}-\hat{r}_{i,i,m}\big|}{r_{i,i,m}}\\
		&=\bO_{p}\{(\log d/n_{m})^{1/2}\}.
		\end{split}
		\end{equation*}
		By Lemma \ref{lemma:inverse} and Assumption \textbf{(B)}, we have
		\begin{equation}\label{equ:M''}
		\max_{1\le i<j\le d}L^{''}_{i,j,m}=\bO_{p}\{(\log d/n_{m})^{1/2}\}.
		\end{equation}
		Hence, combining \eqref{equ:thetaijhat} and the each bound in \eqref{equ:M}, \eqref{equ:M'} and \eqref{equ:M''}, we have
		\begin{equation}\label{equ:maxsigmahat1}
		\max_{1\le s \le d(d-1)/2 \atop m=1,2} |\hat{\sigma}_{s,s,m}-\sigma_{s,s,m}|=\bO_{p}\{(\log d/n_{m})^{1/2}\}.
		\end{equation}

		\textbf{Step 2.} For $s\neq t$, by the definition of $\sigma_{s,t,m}$ and $\hat{\sigma}_{s,t,m}$ in (4.1) and (A.3), we have
		\begin{equation*}
		|\hat{\sigma}_{s,t,m}-\sigma_{s,t,m}|= \bigg|\frac{\hat{r}_{i_{1},i_{2},m}\hat{r}_{j_{1},j_{2},m}+\hat{r}_{i_{1 }, j_{2},m}\hat{r}_{i_{2},j_{1},m}}{n_m\hat{r}_{i_1,i_1,m}\hat{r}_{j_1,j_1,m}\hat{r}_{i_2,i_2,m}\hat{r}_{j_2,j_2,m}}-	
		\frac{r_{i_{1},i_{2},m}r_{j_{1},j_{2},m}\!+\!r_{i_{1},j_{2},m}r_{i_{2},j_{1},m}}{n_mr_{i_1,i_1,m}r_{j_1,j_1,m}r_{i_2,i_2,m}r_{j_2,j_2,m}}\bigg|.
		\end{equation*}
		By the triangle inequality, we have
		\begin{equation}\label{equ:D1D2}
		|\hat{\sigma}_{s,t,m}-\sigma_{s,t,m}|\le D_1D_2,
		\end{equation}
		where  $D_2=\bigg|\frac{r_{i_1,i_1,m}r_{j_1,j_1,m}r_{i_2,i_2,m}r_{j_2,j_2,m}}{\hat{r}_{i_1,i_1,m}\hat{r}_{j_1,j_1,m}\hat{r}_{i_2,i_2,m}\hat{r}_{j_2,j_2,m}}-1\bigg|$ and $D_1=L_1+L_2$ with
		\begin{equation*}
		L_1=\frac{\big|r_{i_{1},i_{2},m}r_{j_{1},j_{2},m}\!-\hat{r}_{i_{1 }, i_{2},m}\hat{r}_{j_{1},j_{2},m}\big|
		}{n_mr_{i_1,i_1,m}r_{j_1,j_1,m}r_{i_2,i_2,m}r_{j_2,j_2,m}},  \qquad
		L_2=\frac{\big|r_{i_{1},j_{2},m}r_{i_{2},j_{1},m}\!-\hat{r}_{i_{1 }, j_{2},m}\hat{r}_{i_{2},j_{1},m}\big|
		}{n_mr_{i_1,i_1,m}r_{j_1,j_1,m}r_{i_2,i_2,m}r_{j_2,j_2,m}}.
		\end{equation*}
		To bound $D_1$, we bound $L_1$ and $L_2$ separately. For $L_1$, by triangle inequality, we have
		\begin{equation*}
		\begin{split}
		L_1 \le&\frac{\big|(r_{i_{1},i_{2},m}-\hat{r}_{i_{1 }, i_{2},m})(r_{j_{1},j_{2},m}-\hat{r}_{j_{1},j_{2},m})\big|+\big|r_{j_{1},j_{2},m}(r_{i_{1},i_{2},m}\!-\!\hat{r}_{i_{1},i_{2},m})\big|
		}{n_mr_{i_1,i_1,m}r_{j_1,j_1,m}r_{i_2,i_2,m}r_{j_2,j_2,m}}\\
		&+\frac{\big|(\hat{r}_{j_{1},j_{2},m}-r_{j_{1},j_{2},m})(r_{i_{1},i_{2},m}\!-\!\hat{r}_{i_{1},i_{2},m})\big|+\big|r_{i_{1},j_{2},m}(r_{j_{1},j_{2},m}\!-\!\hat{r}_{j_{1},j_{2},m})\big|}{n_mr_{i_1,i_1,m}r_{j_1,j_1,m}r_{i_2,i_2,m}r_{j_2,j_2,m}}
		\\
		&+\frac{\big|(\hat{r}_{i_{1},j_{2},m}-r_{i_{1},j_{2},m})(r_{j_{1},j_{2},m}\!-\!\hat{r}_{j_{1},j_{2},m})\big|}{n_mr_{i_1,i_1,m}r_{j_1,j_1,m}r_{i_2,i_2,m}r_{j_2,j_2,m}}.
		\end{split}
		\end{equation*}
		By Lemma A.3 and Assumption \textbf{(B)}, we obtain
		\begin{equation}\label{equ:DL1}
		\max_{(i_1,j_1)\neq(i_2j_2)\atop 1\le i_1<j_1\le d, 1\le i_2<j_2\le d} L_1 = \bO_{p}\{(\log d/n_{m})^{1/2}\}.
		\end{equation}
		With similar arguments, we have
		\begin{equation}\label{equ:DL2}
		\max_{(i_1,j_1)\neq(i_2j_2)\atop 1\le i_1<j_1\le d, 1\le i_2<j_2\le d} L_2 = \bO_{p}\{(\log d/n_m)^{1/2}\}.
		\end{equation}
		Combining \eqref{equ:DL1} and \eqref{equ:DL2}, we get
		\begin{equation}\label{equ:D1le}
		\max_{(i_1,j_1)\neq(i_2j_2)\atop 1\le i_1<j_1\le d, 1\le i_2<j_2\le d} D_1 =  \bO_{p}\{(\log d/n)^{1/2}\}
		\end{equation}
		By Lemma \ref{lemma:inverse}, to bound $D_2$, we only need to bound $\bigg|\frac{\hat{r}_{i_1,i_1,m}\hat{r}_{j_1,j_1,m}\hat{r}_{i_2,i_2,m}\hat{r}_{j_2,j_2,m}}{r_{i_1,i_1,m}r_{j_1,j_1,m}r_{i_2,i_2,m}r_{j_2,j_2,m}}-1\bigg|$.
		By the triangle inequality and Lemma A.3, we obatain
		\begin{equation}\label{equ:D2le}
		\max\limits_{(i_1,j_1)\neq(i_2j_2)\atop 1\le i_1<j_1\le d, 1\le i_2<j_2\le d} D_2=	\bO_{p}\{(\log d/n_{m})^{1/2}\}.
		\end{equation}
		Combining \eqref{equ:D1D2}, \eqref{equ:D1le} and \eqref{equ:D2le}, we have
		\begin{equation}\label{equ:maxsigmahat2}
		\max \limits_{1\le s<t \le d(d-1)/2\atop m=1,2} \big|\hat{\sigma}^{\tilde{U}}_{s,t,m}-\sigma^{\tilde{U}}_{s,t,m}\big|=
		\bO_{p}\{(\log d/n)\}.
		\end{equation}
		Hence, by combining \eqref{equ:maxsigmahat1} and\eqref{equ:maxsigmahat2}, we get
		\begin{equation}\label{equ:maxsigmahat}
		\max \limits_{1\le s<t \le d(d-1)/2\atop m=1,2} \big|\hat{\sigma}^{\tilde{U}}_{s,t,m}-\sigma^{\tilde{U}}_{s,t,m}\big|=
		\bO_{p}\{(\log d/n)\}.
		\end{equation}
		
		By the similar argument of the proof of Lemma A.6 in \cite{Zhou2018Unified} and the results in \eqref{equ:maxsigmahat}, we have
		\begin{equation*}
		\max \limits_{1\le s,t \le d(d-1)/2\atop m=1,2} \big|\hat{r}^{\tilde{U}}_{s,t,m}-r^{\tilde{U}}_{s,t,m}\big|=
		\bO_{p}\{(\log d/n)^{1/2}\}.
		\end{equation*}
		Hence, the proof of Lemma A.4 is finished.
	\end{proof}
	
	\subsection{Proof of Lemma A.5}
	\begin{proof}
		The proof procedure for $s\neq t$ is similar to $s=t$. Hence, to illustrate the proof, we only show the proof for $\tilde{\theta}_{i,j,m}={\hat{\sigma}^{\bT^{b}}_{s,t,m}}$ with $s=t$.
		By the definition of $T^{b}_{i,j,m}$ and $\hat{r}_{i,j,m}$ in (3.3) and (2.5) respectively we obtain,
		\begin{equation*}\label{tildethetaest}
		\begin{split}
		\tilde{\theta}_{i,j,m} \!=\! \frac{1}{r^{2}_{i,i,m}r^{2}_{j,j,m}n_{m}}\underbrace{\frac{1}{n_{m}}\sum_{k=1}^{n_{m}}\Big[\big(\hat{\epsilon}_{k,i,m}\hat{\epsilon}_{k,j,m}\!+\!\hat{\beta}_{i,j,m}\hat{\epsilon}^{2}_{k,i,m}+\hat{\beta}_{j\!-\!1,i,m}\hat{\epsilon}^{2}_{k,j,m})\!+\!\hat{r}_{i,j,m}\Big]^{2}}_{L_1}\underbrace{
			\Big|\frac{r^{2}_{i,i,m}r^{2}_{j,j,m}}{\hat{r}^{2}_{i,i,m}\hat{r}^{2}_{j,j,m}}\Big|}_{L_2}.
		\end{split}
		\end{equation*}
		where $1\le i<j\le d$. By Lemma \ref{lemma:inverse} and  combining (2.6), we have $L_{2}=1+\bO_p\{(\log d/n_{m})^{1/2}\}$.
		
		To bound $L_1$, we introduce another notation first. Let $\tilde{\epsilon}_{k,i,m}=\epsilon_{k,i,m}-\bar{\epsilon}_{i,m}$, where $\bar{\epsilon}_{i,m}=1/n_{m}\sum_{k=1}^{n_m}\epsilon_{k,i,m}$. Hence,  by inserting $\tilde{\epsilon}_{k,i,m}\tilde{\epsilon}_{k,j,m}$, $L_{1}$ can be rewritten as:
		\begin{equation*}
		\begin{split}
		L_{1}=&\underbrace{\frac{1}{n_{m}}\sum_{k=1}^{n_{m}}\Big[\big(\hat{\epsilon}_{k,i,m}\hat{\epsilon}_{k,j,m}\!+\!\hat{\beta}_{i,j,m}\hat{\epsilon}^{2}_{k,i,m}+\hat{\beta}_{j\!-\!1,i,m}\hat{\epsilon}^{2}_{k,j,m})\!+\!\tilde{\epsilon}_{k,i,m}\tilde{\epsilon}_{k,j,m}\Big]^{2}}_{L_{11}}\!+\!\underbrace{\frac{1}{n_{m}}\sum_{k=1}^{n_{m}}(\tilde{\epsilon}_{k,i,m}\tilde{\epsilon}_{k,j,m}\!-\!\hat{r}_{i,j,m})^{2}}_{L_{12}}\\
		&+\underbrace{\frac{1}{n_{m}}\sum_{k=1}^{n_{m}}\Big[\big(\hat{\epsilon}_{k,i,m}\hat{\epsilon}_{k,j,m}\!+\!\hat{\epsilon}^{2}_{k,i,m}\hat{\beta}_{i,j,m}+\hat{\epsilon}^{2}_{k,j,m}\hat{\beta}_{j\!-\!1,i,m})\!+\!\tilde{\epsilon}_{k,i,m}\tilde{\epsilon}_{k,j,m}\Big](\tilde{\epsilon}_{k,i,m}\tilde{\epsilon}_{k,j,m}-\hat{r}_{i,j,m})}_{L_{13}}.
		\end{split}
		\end{equation*}
		
		{By the decomposition of $L_{1}$, by Cauchy-Schwarz inequality, we have that $L_{13}\le \sqrt{L_{11}L_{12}}$. Hence, we only need to bound $L_{11}$ and $L_{12}$. In the following, we  obtain the bounds for $L_{11}$, $L_{12}$ respectively.}
		By the model setting in (2.1) and by the definition of $\hat{\epsilon}_{k,i,m}$ in (2.3), we have,
		\begin{equation*}
		\hat{\epsilon}_{k,i,m}=\tilde{\epsilon}_{k,i,m}-(\bX_{k,-i,m}-\bar{\bX}_{-i,m})(\hat{\bbeta}_{i,m}-\bbeta_{i,m}).
		\end{equation*}
		{Thus,  with these new notations, we decompose the main term $\hat{\epsilon}_{k,i,m}\hat{\epsilon}_{k,j,m}$ of $L_{11}$ and $L_{13}$ as following. For $1\le i<j\le d$, we obtain}
		\begin{equation}\label{hatepsilonij}
		\begin{split}
		\hat{\epsilon}_{k,i,m}\hat{\epsilon}_{k,j,m}=&\tilde{\epsilon}_{k,i,m}\tilde{\epsilon}_{k,j,m}-
		\tilde{\epsilon}_{k,i,m}(\bX_{k,-j,m}-\bar{\bX}_{-j,m})(\hat{\bbeta}_{j,m}-\bbeta_{j,m})\\
		&-\tilde{\epsilon}_{k,j,m}(\bX_{k,-i,m}-\bar{\bX}_{-i,m})(\hat{\bbeta}_{i,m}-\bbeta_{i,m})\\
		&+(\hat{\bbeta}_{i,m}-\bbeta_{i,m})^{\top}(\bX_{k,-i,m}-\bar{\bX}_{-i,m})^{\top}(\bX_{k,-j,m}-\bar{\bX}_{-j,m})(\hat{\bbeta}_{j,m}-\bbeta_{j,m}).
		\end{split}
		\end{equation}
		For $1\le i=j\le d$, we obtain
		\begin{equation}\label{hatepsiloni2}
		\begin{split}
		\hat{\epsilon}^{2}_{k,i,m}=&\tilde{\epsilon}^{2}_{k,i,m}-
		2\tilde{\epsilon}_{k,i,m}(\bX_{k,-i,m}-\bar{\bX}_{-i,m})(\hat{\bbeta}_{i,m}-\bbeta_{i,m})\\
		&+(\hat{\bbeta}_{i,m}-\bbeta_{i,m})^{\top}(\bX_{k,-i,m}-\bar{\bX}_{-i,m})^{\top}(\bX_{k,-i,m}-\bar{\bX}_{-i,m})(\hat{\bbeta}_{i,m}-\bbeta_{i,m}).
		\end{split}
		\end{equation}
		{With these new notations and together with \eqref{hatepsilonij} and \eqref{hatepsiloni2}, we decompose $L_{11}$ as}
\[
		\begin{array}{llllllll}
		L_{11}=&\frac{1}{n_{m}}\sum_{k=1}^{n_{m}}\Big[
		\underbrace{(2\tilde{\epsilon}_{k,i,m}\tilde{\epsilon}_{k,j,m}\!+\!\beta_{i,j,m}\tilde{\epsilon}^{2}_{k,i,m}+\beta_{j\!-\!1,i,m}\tilde{\epsilon}^{2}_{k,j,m})}_{L_{11,1,ijk}}\\
		&+\underbrace{(\hat{\bbeta}_{i,m}-\bbeta_{i,m})^{\top}(\bX_{k,-i,m}-\bar{\bX}_{-i,m})^{\top}(\bX_{k,-j,m}-\bar{\bX}_{-j,m})(\hat{\bbeta}_{j,m}-\bbeta_{j,m})}_{L_{11,2,ijk}}\\
		&-\underbrace{\tilde{\epsilon}_{k,i,m}(\bX_{k,-j,m}-\bar{\bX}_{-j,m})^{\top}(\hat{\bbeta}_{j,m}-\bbeta_{j,m})}_{L_{11,3,ijk}}-
		\underbrace{\tilde{\epsilon}_{k,j,m}(\bX_{k,-i,m}-\bar{\bX}_{-i,m})^{\top}(\hat{\bbeta}_{i,m}-\bbeta_{i,m})}_{L_{11,4,ijk}}\\
\end{array}
\]
\[	\begin{array}{llllllll}
&\!-\!\underbrace{2\beta_{i,j,m}\tilde{\epsilon}_{k,i,m}(\bX_{k,\!-\!i,m}\!-\!\bar{\bX}_{\!-\!i,m})(\hat{\bbeta}_{i,m}\!-\!\bbeta_{i,m})}_{L_{11,5,ijk}}\!-\!\underbrace{2\beta_{j\!-\!1,i,m}\tilde{\epsilon}_{k,j,m}(\bX_{k,\!-\!j,m}\!-\!\bar{\bX}_{k,\!-\!j,m})^{\top}(\hat{\bbeta}_{j,m}\!-\!\bbeta_{j,m})}_{L_{11,6,ijk}}\\
		&+\underbrace{\beta_{i,j,m}(\hat{\bbeta}_{i,m}-\bbeta_{i,m})^{\top}(\bX_{k,-i,m}-\bar{\bX}_{-i,m})^{\top}(\bX_{k,-i,m}-\bar{\bX}_{-i,m})(\hat{\bbeta}_{i,m}-\bbeta_{i,m})}_{L_{11,7,ijk}}\\
		&+\underbrace{\beta_{j-1,i,m}(\hat{\bbeta}_{j,m}-\bbeta_{j,m})^{\top}(\bX_{k,-j,m}-\bar{\bX}_{-j,m})^{\top}(\bX_{k,-j,m}-\bar{\bX}_{-j,m})(\hat{\bbeta}_{j,m}-\bbeta_{j,m})}_{L_{11,8,ijk}}\\
		&+\underbrace{(\hat{\beta}_{i,j,m}-\beta_{i,j,m})\tilde{\epsilon}_{k,i,m}^{2}}_{L_{11,9,ijk}}-\underbrace{2(\hat{\beta}_{i,j,m}-\beta_{i,j,m})\tilde{\epsilon}_{k,i,m}(\bX_{k,-i,m}-\bar{\bX}_{-i,m})(\hat{\bbeta}_{i,m}-\bbeta_{i,m})^{\top}}_{L_{11,10,ijk}}\\
		&+\underbrace{(\hat{\beta}_{i,j,m}-\beta_{i,j,m})(\hat{\bbeta}_{i,m}-\bbeta_{i,m})^{\top}(\bX_{k,-i,m}-\bar{\bX}_{-i,m})^{\top}(\bX_{k,-i,m}-\bar{\bX}_{-i,m})(\hat{\bbeta}_{i,m}-\bbeta_{i,m})}_{L_{11,11,ijk}}\\
		&+\underbrace{(\hat{\beta}_{j\!-\!1,i,m}\!-\!\beta_{j\!-\!1,i,m})\tilde{\epsilon}_{k,j,m}^{2}}_{L_{11,12,ijk}}\!-\!\underbrace{2(\hat{\beta}_{j\!-\!1,i,m}\!-\!\beta_{j\!-\!1,i,m})\tilde{\epsilon}_{k,j,m}(\bX_{k,\!-\!j,m}\!-\!\bar{\bX}_{k,\!-\!j,m})^{\top}(\hat{\bbeta}_{j,m}\!-\!\bbeta_{j,m})}_{L_{11,13,ijk}}\\
		&\!+\!\underbrace{(\hat{\beta}_{j\!-\!1,i,m}\!-\!\beta_{j\!-\!1,i,m})(\hat{\bbeta}_{j,m}\!-\!\bbeta_{j,m})^{\top}(\bX_{k,\!-\!j,m}\!-\!\bar{\bX}_{\!-\!j,m})^{\top}(\bX_{k,\!-\!j,m}\!-\!\bar{\bX}_{\!-\!j,m})(\hat{\bbeta}_{j,m}\!-\!\bbeta_{j,m})}_{L_{11,14,ijk}}\Big]^2.
\end{array}
\]
		By the triangle inequality $(\sum_{k=1}^{K}a_{k})^2\le K\sum_{k=1}^{K}a_{k}^{2}$, we have
		\begin{equation}\label{equ:L11}
		L_{11}\le 14\Big[\frac{1}{n_{m}}\sum_{k=1}^{n_{m}}L_{11,1,ijk}^2+\ldots+\frac{1}{n_{m}}\sum_{k=1}^{n_{m}}L_{11,14,ijk}^2\Big].
		\end{equation}
		{In the following, we control each component respectively.}
		For $\epsilon_{k,i,m}$ with $i=1,\ldots,d$ are zero mean guassian distribution random variables, i.e. $\epsilon_{k,i,m}$ with $i=1,\ldots,d$ are sub-guassian random variables. By Lemma A.2, $\epsilon_{k,i,m}^{2}$ or $\epsilon_{k,i,m}\epsilon_{k,j,m}$ are following sub-exponential distribution. Further, $L_{11,1}$ is a sub-exponential random variable. Accroding to Theorem 6 of \cite{Delaigle2011Robustness}, we could have
		\begin{equation}\label{equ:L111ijk}
		\max_{1\le i<j\le d}\Big|\frac{1}{n_{m}}\sum_{k=1}^{n_{m}}
		L^{2}_{11,1,ijk}\Big| \le C \sqrt{\frac{\log(dn_{m})}{n_{m}}}+C_{1}\frac{\log^{2}(dn_{m})}{n_{m}}.
		\end{equation}
		{For any $i$, $j$ and $k$, $L_{11,2,ijk}$ is nonnegative. Hence, we have $a^2+b^2\le (a+b)^2$ and }
		\begin{equation*}
		\frac{1}{n_{m}}\sum_{k=1}^{n_{m}}L_{11,2,ijk}^2\le n_{m}\bigg[\frac{1}{n_{m}}\sum_{k=1}^{n_{m}}L_{11,2,ijk}\bigg]^{2}.
		\end{equation*}
		By the triangle inequality, we have,
		\begin{equation*}
		\begin{split}
		\frac{1}{n_{m}}\sum_{k=1}^{n_{m}}L_{11,2,ijk}=&\Big|(\hat{\bbeta}_{i,m}-\bbeta_{i,m})^{\top}\widehat{\bSigma}_{-i,-j,m}(\hat{\bbeta}_{j,m}-\bbeta_{j,m})\Big| \le  \Big|(\hat{\bbeta}_{i,m}-\bbeta_{i,m})^{\top}\bSigma_{-i,-j,m}(\hat{\bbeta}_{j,m}-\bbeta_{j,m})\Big|\\
		&+\Big|(\hat{\bbeta}_{i,m}-\bbeta_{i,m})^{\top}(\widehat{\bSigma}_{-i,-j,m}-\bSigma_{-i,-j,m})(\hat{\bbeta}_{j,m}-\bbeta_{j,m})\Big|.
		\end{split}
		\end{equation*}
		It's easy to show that there exists an constant $C>0$ such that for any $M>0$,
		\begin{equation}\label{Cai2011}
		\P\Big(\max_{1\le i \le j\le d}\big|\hat{\sigma}_{i,j,m}-\sigma_{i,j,m}\big|\ge C\sqrt{\log d/n_{m}}\Big)=\bO\Big(d^{-M}\Big).
		\end{equation}
		By Assumption \textbf{(B)} and \textbf{(D)}, the Lasso estimator of $\beta_{i,m}$ satisfied the bound in (2.2). Furthermore, we obtain,
		\begin{equation*}
		\max_{1\le i <j\le d} \Big|(\hat{\bbeta}_{i,m}-\bbeta_{i,m})^{\top}(\widehat{\bSigma}_{-i,-j,m}-\bSigma_{-i,-j,m})(\hat{\bbeta}_{j,m}-\bbeta_{j,m})\Big|=\bop\Big(n_{m}^{-1/2}(\log d)^{-3/2}\Big).
		\end{equation*}
		{By Assumption \textbf{(B)}, we have $\lambda_{\max}(\bSigma_{m})\le C$. Further, by Cauchy-Schwarz inequality and the conditions in (2.2), we have}
		\begin{equation*}
		\begin{split}
		\max_{1\le i <j\le d}\Big|(\hat{\bbeta}_{i,m}-\bbeta_{i,m})^{\top}\bSigma_{-i,-j,m}(\hat{\bbeta}_{j,m}-\bbeta_{j,m})\Big|&=\big\|\bSigma_{-i,-j,m}\big\|\max_{1\le i \le d}\big|\hat{\bbeta}_{i,m}-\bbeta_{i,m}\big|\max_{1\le j \le d}\big|\hat{\bbeta}_{j,m}-\bbeta_{j,m}\big|\\
		&=\bop\big\{(n_{m}\log d)^{-1/2}\big\}.
		\end{split}
		\end{equation*}
		{Hence, combining these results, we have}
		\begin{equation}\label{equ:L112ijk}
		\max_{1\le i<j\le d}\frac{1}{n_{m}}\sum_{k=1}^{n_{m}}L_{11,2,ijk}^2=\bop\Big\{(\log d)^{-1}\Big\}.
		\end{equation}
		{For $1/n_{m}\sum_{k=1}^{n_{m}}L^{2}_{11,3,ijk}$, by Cauchy–Schwarz inequality and the triangle inequality, we have}
		\begin{equation*}
		\begin{split}
		\frac{1}{n_{m}}&\sum_{k=1}^{n_{m}}L^{2}_{11,3,ijk}\!=\!\frac{1}{n_{m}}\sum_{k=1}^{n_{m}}\tilde{\epsilon}^{2}_{k,i,m}(\hat{\bbeta}_{j,m}\!-\!\bbeta_{j,m})^{\top}(\bX_{k,-j,m}\!-\!\bar{\bX}_{-j,m})(\bX_{k,-j,m}\!-\!\bar{\bX}_{-j,m})^{\top}(\hat{\bbeta}_{j,m}-\bbeta_{j,m})\\
		& \!\le\! \sqrt{\frac{1}{n_{m}}\sum_{k=1}^{n_{m}}{\tilde{\epsilon}^{4}_{k,i,m}}n_{m}\frac{1}{n^{2}_{m}}\sum_{k=1}^{n_{m}}\big[(\hat{\bbeta}_{j,m}\!-\!\bbeta_{j,m})^{\top}(\bX_{k,-j,m}\!-\!\bar{\bX}_{-j,m})(\bX_{k,-j,m}\!-\!\bar{\bX}_{-j,m})^{\top}(\hat{\bbeta}_{j,m}\!-\!\bbeta_{j,m})\big]^{2}}\\
		& \!\le\! \sqrt{n_{m}}\sqrt{\frac{1}{n_{m}}\sum_{k=1}^{n_{m}}{\tilde{\epsilon}^{4}_{k,i,m}}}
		\underbrace{
			\Bigg[\frac{1}{n_{m}}\sum_{k=1}^{n_{m}}(\hat{\bbeta}_{j,m}\!-\!\bbeta_{j,m})^{\top}(\bX_{k,-j,m}\!-\!\bar{\bX}_{\!-\!j,m})(\bX_{k,\!-\!j,m}\!-\!\bar{\bX}_{\!-\!j,m})^{\top}(\hat{\bbeta}_{j,m}\!-\!\bbeta_{j,m})\Bigg]}_{L^{'}_{11,3}}.
		\end{split}
		\end{equation*}
		{By similar arguments of the bounding process of $1/n_{m}\sum_{k=1}^{n_{m}}L_{11,2,ijk}$, we have $$\max_{1\le j\le d}L^{'}_{11,3}=\bop\Big\{(n_{m}\log d)^{-1/2}\Big\}.$$ Besides, $\tilde{\epsilon}_{k,i,m}$ follows zero mean Gaussian distribution. By Lemma \ref{lemma:Isserlistheorem}, and by Assumption \textbf{(B)}, we obtain $E[\tilde{\epsilon}^{4}_{k,i,m}]\le C$, where $C$ is some positive constant. And, by Lemma A.2, $\tilde{\epsilon}^{2}_{k,i,m}$ follows sub-exponential distribution. Moreover, according to Theorem 6 in \cite{Delaigle2011Robustness}, we have
		}
		\begin{equation*}
		\max_{1\le i\le d}\Big|\frac{1}{n_{m}}\sum_{k=1}^{n_{m}}{\tilde{\epsilon}^{4}_{k,i,m}}\Big|\le 3c_0+\sqrt{\frac{\log(dn_{m})}{n_{m}}}+C_{1}\frac{\log^{2}(dn_{m})}{n_{m}}.
		\end{equation*}
		{Hence, we  have}
		\begin{equation}\label{equ:L113ijk}
		\max_{1\le i<j\le d}\frac{1}{n_{m}^2}\sum_{k=1}^{n_{m}}L_{11,3,ijk}^2=\bop\Big\{(\log d)^{-1/2}\Big\}.
		\end{equation}
		{By similar arguments, we have }
		\begin{equation*}\label{equ:L1148ijk}
		\max_{1\le i<j\le d}\frac{1}{n_{m}^2}\sum_{k=1}^{n_{m}}L_{11,q,ijk}^2= \left\{\
		\begin{aligned}
		&\bop\Big\{(\log d)^{-1/2}\Big\}, \qquad \quad q=4,5,6,7,8,\\
		&\bop\Big\{(n_{m}\log d)^{-1/2}\Big\}, \qquad q=9,12,\\
		&\bop\Big\{n^{-1/2}_m(\log d)^{-1}\Big\}, \qquad q=10,11,13,14.
		\end{aligned}
		\right.
		\end{equation*}
		Further, combining these results and \eqref{equ:L11}, \eqref{equ:L111ijk}, \eqref{equ:L112ijk}, \eqref{equ:L113ijk}, we could get
		\begin{equation}\label{equ:L11bound}
		L_{11}=\bop\Big\{(\log d)^{-1/2}\Big\}.
		\end{equation}

		For $L_{12}$, by triangle inequality, we could rewritten it as:
		\begin{equation*}
		\begin{split}
		L_{12}=&\underbrace{\frac{1}{n_{m}}\sum_{k=1}^{n_{m}}\big(\epsilon_{k,i,m}\epsilon_{k,j,m}-\E[\epsilon_{k,i,m}\epsilon_{k,j,m}]\big)^{2}}_{L_{12,1,ij}}
		+\underbrace{\frac{1}{n_{m}}\big(\E[\epsilon_{k,i,m}\epsilon_{k,j,m}]-\hat{r}_{i,j,m}\big)^{2}}_{L_{12,2,ij}}\\
		&+\underbrace{\frac{1}{n_{m}}\sum_{k=1}^{n_{m}}\big(\epsilon_{k,i,m}\epsilon_{k,j,m}-\E[\epsilon_{k,i,m}\epsilon_{k,j,m}]\big)\big(\E[\epsilon_{k,i,m}\epsilon_{k,j,m}]-\hat{r}_{i,j,m}\big)}_{L_{12,3,ij}},
		\end{split}
		\end{equation*}
		With similiar arguments of $L_{11,1,ijk}$, and by  Theorem 6 \cite{Delaigle2011Robustness} and get
		\begin{equation*}
		\max_{1\le i<j\le d}\Big|L_{12,1,ij}-\text{Var}(\epsilon_{k,i,m}\epsilon_{k,j,m})\Big| \le C\sqrt{\frac{\log(dn_{m})}{n_{m}}}+C_{1}\frac{\log^{2}(dn_{m})}{n_{m}},
		\end{equation*}
		By Lemma A.3, we could get $\max_{1\le i<j\le d}L_{12,2,ij}=\bop(\log d/n^{2}_{m})$. By the Cauchy-Schwarz inequality, we have $\max_{1\le i<j\le d}L_{12,3,ij}=\bop(\log d/n^{2}_{m})$.
		Hence, we could get
		\begin{equation}\label{equ:L12bound}
		L_{12}=\text{Var}(\epsilon_{k,i,m}\epsilon_{k,j,m})+\bop(\log d/n^{2}_{m}).
		\end{equation}
		Combining the bound of $L_{11}$ in \eqref{equ:L11bound}, we obtain $L_{13}=\bop\{(\log d)^{-1/4}\}$. Furthermore, Combining each bound for $L_{11}$, $L_{12}$ and $L_{13}$, we obtain $L_{1}=\text{Var}(\epsilon_{k,i,m}\epsilon_{k,j,m})+\bop\{(\log d)^{-1/4}\}$. Hence, we could get
		\begin{equation*}
		\max_{1\le i<j\le d}\big|\tilde{\theta}_{i,j,m}-\theta_{i,j,m}\big|=\bOp\{(\log d/n_{m})^{1/2}\}.
		\end{equation*}
		With similar argument, we get
		\begin{equation}\label{equ:simatt}
		\max_{1\le s,t \le d(d-1)/2\atop m=1,2}\big|\hat{\sigma}^{\Tb^{b}}_{s,t,m}-\sigma^{\tilde{U}}_{s,t,m}\big|=\bOp\{(\log d/n_{m})^{1/2}\}.
		\end{equation}
		By the similar argument of the proof of Lemma A.6 in \cite{Zhou2018Unified} and in \eqref{equ:simatt}, we have
		\begin{equation*}
		\max_{1\le s,t \le d(d-1)/2\atop m=1,2}\big|\hat{r}^{\Tb^{b}}_{s,t,m}-r^{\tilde{U}}_{s,t,m}\big|=\bOp\{(\log d/n_{m})^{1/2}\}.
		\end{equation*}
		Hence, the proof of Lemma A.5 is finished.
		
	\end{proof}

	\subsection{Proof of Lemma B.1}\label{proof:lemmaB.1}
	\begin{proof}
		To bounded $\|{\rm trivec}(\Wb)-{\rm trivec}(\Hb)\|_{(s_0,p)}$, we introduce some other notations. Define $\Hb^{*} \in \reals^{d\times d}$ with
		\begin{equation}\label{def:hatH}
		H^{*}_{i,j}=(T_{i,j,1}-T_{i,j,2})/\sqrt{\theta_{i,j,1}+\theta_{i,j,2}}, \qquad  1\le i,j\le d.
		\end{equation}
		By the triangle equity, we have
		\begin{equation}\label{equ:WH}
		\|{\rm trivec}(\Wb)-{\rm trivec}(\Hb)\|_{(s_0,p)} \le \|{\rm trivec}(\Wb)-{\rm trivec}(\Hb^{*})\|_{(s_0,p)}+\|{\rm trivec}(\Hb)-{\rm trivec}(\Hb^{*})\|_{(s_0,p)}.
		\end{equation}
		By the definition of $(s_0,p)$-norm, we have $\|{\rm trivec}(\Wb)-{\rm trivec}(\Hb^{*})\|_{(s_0,p)}\le s_0^{1/p}\|{\rm trivec}(\Wb)-{\rm trivec}(\Hb^{*})\|_{\infty}$.
		Let $\Wb=(W_{i,j})_{d\times d}$ and by the definition of $W_{i,j}$ and $\Hb^{*}$ in (2.6) and \eqref{def:hatH}, we have $\|{\rm trivec}(\Wb)-{\rm trivec}(\Hb^{*})\|_{\infty}\le L_1 L_2$, where
		\begin{equation*}
		L_1=\max_{1\le i < j\le d}\frac{|T_{i,j,1}-T_{i,j,2}|}{\sqrt{\theta_{i,j,1}+\theta_{i,j,2}}} \qquad
		L_2=\max_{1\le i <j \le d}\Big|1-\frac{\sqrt{\theta_{i,j,1}+\theta_{i,j,2}}}{\sqrt{\hat{\theta}_{i,j1}+\hat{\theta}_{i,j,2}}}\Big|.
		\end{equation*}
		We then analyze $L_1$ and $L_2$ separately. For $L_1$, by triangle inequality, we have
		\begin{equation}\label{equ:L1}
		L_1\le \max_{1\le i < j\le d}\frac{|T_{i,j,1}-\tilde{U}_{i,j,1}|+|T_{i,j,2}-\tilde{U}_{i,j,2}|+|\tilde{U}_{i,j,1}-\tilde{U}_{i,j,2}|}{\sqrt{\theta_{i,j,1}+\theta_{i,j,2}}}.
		\end{equation}  		
		As the Lemma A.2 of \cite{Xia2015Testing} showed,
		\begin{equation*}
		|T_{i,j,m}-\tilde{U}_{i,j,m}|=\bO_{p}\{(\log d/n_{m})^{1/2}\}r_{i,j,m}+\bo_{p}\{(n_m\log d)^{-1/2}\},
		\end{equation*}
		uniformly for $1\le i<j\le d$. Hence,  we obtain
		\begin{equation}\label{equ:maxTU}
		\max_{1\le i<j\le d}|T_{i,j,m}-\tilde{U}_{i,j,m}|=\bO_{p}\Big\{(\log d/n_{m})^{1/2}\Big\}.
		\end{equation}
		By Lemma A.3, we have
		\begin{equation*}
		\max_{1\le i <j\le d} U_{i,j,m}=\bO_{p}\big\{(\log d/n)^{1/2}\big\}.
		\end{equation*}
		By the definition of $\tilde{U}_{i,j,m}$ and the triangle inequality, we have
		\begin{equation}\label{equ:U12}
		\begin{split}
		\max_{1\le i<j\le d}|\tilde{U}_{i,j,1}-\tilde{U}_{i,j,2}|\le &\bigg|\frac{r_{i,j,1}}{r_{i,i,1}r_{j,j,1}}
		-\frac{r_{i,j,2}}{r_{i,i,2}r_{j,j,2}}\bigg|+\bigg|\frac{U_{i,j,1}}{r_{i,i,1}r_{j,j,1}}\bigg|+\bigg|\frac{U_{i,j,2}}{r_{i,i,2}r_{j,j,2}}\bigg|.
		\end{split}
		\end{equation}
		Under $\Hb_0$, by Lemma A.4 and combining \eqref{equ:L1}, \eqref{equ:maxTU} and \eqref{equ:U12}, we have
		\begin{equation}\label{equ:L1eq}
		L_1 = \bO_{p}\big\{(\log d/n)^{1/2}\big\}
		\end{equation} 		
		For $L_2$, motivated by Lemma \ref{lemma:inverse}, we introduce
		\[
		L'_2 =\max_{1\le i <j \le d}\Bigg|1-\frac{\sqrt{\hat{\theta}_{i,j1}+\hat{\theta}_{i,j,2}}}{\sqrt{\theta_{i,j,1}+\theta_{i,j,2}}}\Bigg|,
		\]	
		According to the Equation (2.1), $\bepsilon_{k,m}=(\epsilon_{k,1,m}, \ldots, \epsilon_{k,d,m})^\top$ can be rewritten as a linear combination of some normal distribution $\bX_{k}$ or $\bY_{k}$. Hence, we obtain that $\bepsilon_{k,m}$ follow normal distribution. We have $\theta_{i,j,m}$ are bounded, and denote the upper bound of $\theta_{i,j,m}$ as $Q$.  By the triangle inequality, we obtain
		\[
		\begin{aligned}
		L_2'&\le
		\max_{1\le i<j\le d}{(\theta_{i,j,1}+\theta_{i,j,2})^{-1}}{|\hat{\theta}_{i,j,1}+\hat{\theta}_{i,j,2}-\theta_{i,j,1}-\theta_{i,j,2}|}\\
		&\le \frac{1}{2Q}\Big(\max_{1\le i<j\le d}|\hat{\theta}_{i,j,1}-\theta_{i,j,1}|+\max_{1\le i<j\le d}|\hat{\theta}_{i,j,2}-\theta_{i,j,2}|\Big).
		\end{aligned}
		\]
		Therefore, by the Step 1 of the proof for  Lemma A.3, we have
		\begin{equation*}
		L'_2=\bO_{p}\big\{(\log d/n)^{1/2}\big\}.
		\end{equation*}
		By Lemma \ref{lemma:inverse}, we have that
		\begin{equation}\label{equ:L2eq}
		L_2=\bO_{p}\big\{(\log d/n)^{1/2}\big\}.
		\end{equation}
		Combining the bounds of $L_1$ and $L_2$ in \eqref{equ:L1eq} and \eqref{equ:L2eq}, we get
		\begin{equation}\label{equ:WH*}
		\|{\rm trivec}(\Wb) -{\rm trivec}(\Hb^{*})\|_{(s_0,p)} =\bO_{p}\Big\{ (s_0\log d)/n^{3/4}\Big\}
		\end{equation}
		By the definition of $\Hb$ and $\Hb^{*}$ and the triangle inequality, we have
		\begin{equation*}
		\|{\rm trivec}(\Hb^{*})-{\rm trivec}(\Hb)\|_{\infty}\le \max_{1\le i<j\le d}\bigg| \frac{T_{i,j,1}-\tilde{U}_{i,j,1}}{\sqrt{\theta_{i,j,1}+\theta_{i,j,2}}}\bigg|+\max_{1\le i<j\le d}\bigg| \frac{T_{i,j,2}-\tilde{U}_{i,j,2}}{\sqrt{\theta_{i,j,1}+\theta_{i,j,2}}}\bigg|
		= \bO_{p}\Big\{(\log d/n)^{1/2}\Big\}.
		\end{equation*}
		Considering  $\|{\rm trivec}(\Hb^{*})-{\rm trivec}(\Hb)\|_{(s_0,p)}\le s_0^{1/p}\|{\rm trivec}(\Hb^{*})-{\rm trivec}(\Hb)\|_{\infty}$, we have
		\begin{equation}\label{equ:HH*}
		\|{\rm trivec}(\Hb^{*})-{\rm trivec}(\Hb)\|_{(s_0,p)}=\bO_{p}\Big\{s_0(\log d/n)^{1/2}\Big\}.
		\end{equation}
		Therefore, combining \eqref{equ:WH}, \eqref{equ:WH*} and \eqref{equ:HH*},
		we have $\|{\rm trivec}(\Wb)-{\rm trivec}(\Hb)\|_{(s_0,p)}= \bO_{p}\big\{s_0(\log d/n)^{1/2}\big\}$, which finishes the proof of Lemma B.1.			
	\end{proof}

	\section{Additional simulation results}\label{sec:simulation}
\begin{figure*}[hbpt]
		\centering
		\includegraphics[width=16cm, height=14.5cm]{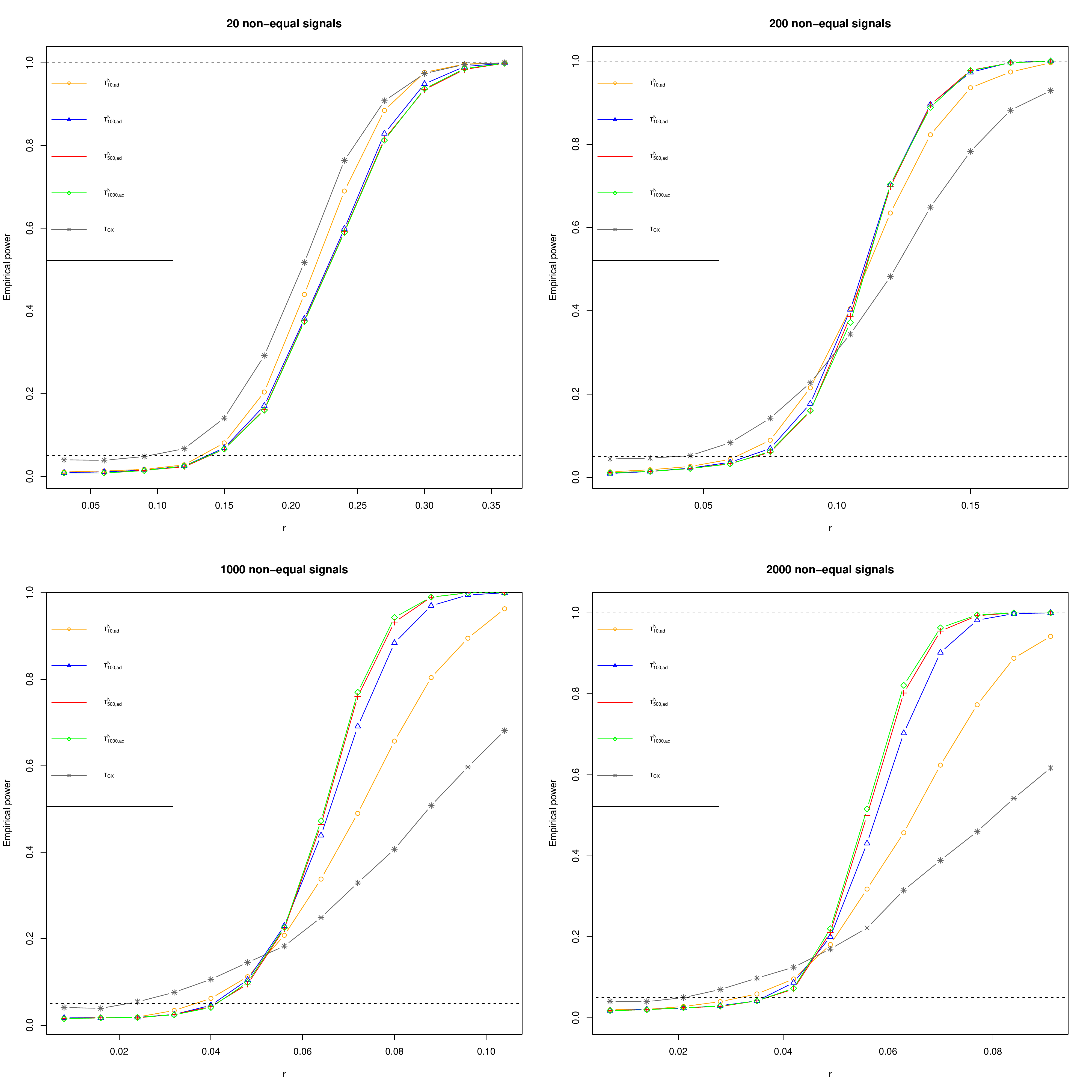}
		\caption{Empirical powers of various tests for {\bf Model 2}. The orange line with circles represents the adaptive test $T^{N}_{10,{\rm ad}}$, the blue line with triangles represents the adaptive test $T^{N}_{100,{\rm ad}}$, the red line with crosses represents the adaptive test $T^{N}_{500,{\rm ad}}$,  the green line with diamonds represents the adaptive test $T^{N}_{1000,{\rm ad}}$, the black line with stars represents the $T_{\rm CX}$ test.$\qquad \qquad \qquad \qquad \qquad \qquad \qquad \qquad \qquad \qquad \qquad \qquad \qquad \qquad \qquad \qquad \qquad$}\label{fig:figmodel2}
	\end{figure*}

\begin{figure*}[hbpt]
		\centering
		\includegraphics[width=16cm, height=16cm]{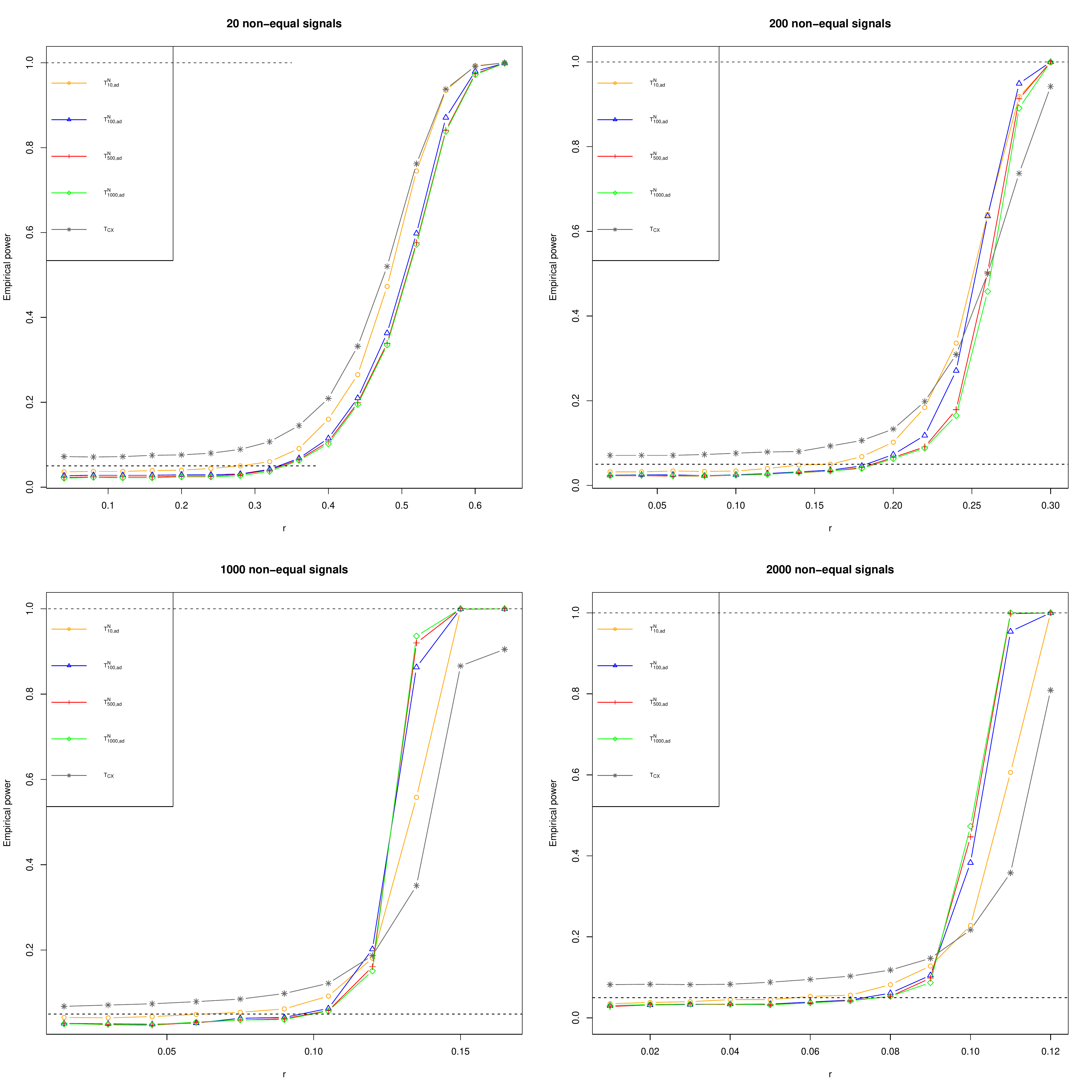}
		\caption{Empirical powers of various tests for {\bf Model 3}. The orange line with circles represents the adaptive test $T^{N}_{10,{\rm ad}}$, the blue line with triangles represents the adaptive test $T^{N}_{100,{\rm ad}}$, the red line with crosses represents the adaptive test $T^{N}_{500,{\rm ad}}$,  the green line with diamonds represents the adaptive test $T^{N}_{1000,{\rm ad}}$, the black line with stars represents the $T_{\rm CX}$ test.$\qquad \qquad \qquad \qquad \qquad \qquad \qquad \qquad \qquad \qquad \qquad \qquad \qquad \qquad \qquad \qquad$}\label{fig:figmodel3}
	\end{figure*}

	In this section, we provide some additional simulation results to illustrate that our method can be adaptive to various models. The empirical performances of the tests for {\bf Model 2} and {\bf Model 3} are shown in Figure \ref{fig:figmodel2} and \ref{fig:figmodel3}.
	The orange line with circles represents the adaptive test $T^{N}_{10,{\rm ad}}$, the blue line with triangles represents the adaptive test $T^{N}_{100,{\rm ad}}$, the red line with crosses represents the adaptive test $T^{N}_{500,{\rm ad}}$,  the green line with diamonds represents the adaptive test $T^{N}_{1000,{\rm ad}}$, the black line with stars represents the $T_{\rm CX}$ test proposed by \cite{Xia2015Testing}. The horizontal axis represents magnitude $r$ in the upper triangle of $\bGamma$,  a larger value of $r$ indicates a stronger signal. The vertical axis represents the empirical powers of different tests, while $r=0$ corresponds to the empirical sizes.

	Similar conclusions for {\bf Model 1} can be drawn for {\bf Model 2} and {\bf Model 3}. Firstly, given the significant level $\alpha=0.05$, the empirical sizes of all the methods for both models are well under control. Secondly, the empirical powers of the maximum test $T_{\rm CX}$ are the highest among  all the tests under the sparse alternative pattern with $m_t=20$. In the meanwhile, the empirical powers of the adaptive test with $s_0=10$ are comparable to those of  $T_{\rm CX}$. Although the empirical powers of the adaptive test grows a bit lower as $s_0$ decreases, the adaptive test with $s_0=1000$ performs comparably. Thirdly, with the numbers of the non-equal elements $m_t$ getting larger, the empirical powers of the adaptive test are getting better and better, especially when $s_0$ increases. Fourthly, the empirical powers of the adaptive tests are not sensitive to  small changes of $s_0$. By Figure \ref{fig:figmodel2} and Figure \ref{fig:figmodel3}, the empirical powers of $T^{N}_{500,{\rm ad}}$ and $T^{N}_{1000,{\rm ad}}$ are almost equal. At last, the adaptive test with $s_0$ close to $m_t/2$ enjoys good performance, for example, the adaptive test with $s_0=10$ is  better than the other adaptive tests for $m_t=20$, the adaptive test $T^{N}_{100,{\rm ad}}$ have a better performance than the other adaptive tests for $m_t=200$, see the top right panel of Figure \ref{fig:figmodel2} and Figure \ref{fig:figmodel3}.

  \end{appendix}


\end{document}